\documentstyle[12pt,aaspp4,flushrt]{article}

\newcommand{\apg}{\gtrsim}
\newcommand{\apl}{\lesssim}
\newcommand{\cmjj}{\mbox{${\rm cm^{-2}}$}}
\newcommand{\etal}{et al.}
\newcommand{\hI}{\mbox{H\,I}}

\newcommand{\ibid}{\underline{\makebox[0.5in]{}}.}

\newcommand{\kms}{\mbox{km\ s${^{-1}}$}}
\newcommand{\lya}{\mbox{${\rm Ly}\alpha$}}

\newcommand{\civ}{\mbox{C\,IV}}
 
\begin{document}
 
\lefthead{Chen et al.}
 
\righthead{}

 
\title{THE GASEOUS EXTENT OF GALAXIES AND THE ORIGIN OF \lya\ ABSORPTION 
SYSTEMS.  V.  OPTICAL AND NEAR-INFRARED PHOTOMETRY OF \lya-ABSORBING GALAXIES 
AT $z < 1$\altaffilmark{1}}
\altaffiltext{1}{Based on observations with the NASA/ESA Hubble Space
Telescope, obtained at the Space Telescope Science Institute, which is operated
by the Association of Universities for Research in Astronomy, Inc., under NASA
contract NAS5--26555.}
 
\author{HSIAO-WEN CHEN\altaffilmark{2,3} and KENNETH M. LANZETTA\altaffilmark{2}}
\affil{Department of Physics and Astronomy, State University of New York at
Stony Brook \\
Stony Brook, NY 11794--3800, U.S.A. \\
lanzetta@sbastr.ess.sunysb.edu}

\altaffiltext{2}{Visiting Astronomer at Infrared Telescope Facility, which is 
operated by the University of Hawaii under contract to the National Aeronautics
and Space Administration}

\altaffiltext{3}{Current address: The Observatories of the Carnegie Institution
of Washington, 813 Santa Barbara Street, Pasadena, CA 91101, U.S.A.  E-Mail: 
hchen@ociw.edu}

\author{JOHN K. WEBB}
\affil{School of Physics, University of New South Wales \\
Sydney 2052, NSW, AUSTRALIA \\
jkw@edwin.phys.unsw.edu.au}

\and

\author{XAVIER BARCONS}
\affil{Instituto de F\'\i sica de Cantabria (Consejo Superior de
Investigaciones Cient\'\i ficas--- \\
Universidad de Cantabria) \\
39005 Santander, SPAIN \\
barcons@ifca.unican.es}

\newpage

\begin{abstract}

  We present results of a program to obtain and analyze HST WFPC2 images and 
ground-based images of galaxies identified in an imaging and spectroscopic 
survey of faint galaxies in fields of HST spectroscopic target QSOs.  
Considering a sample of physically correlated galaxy and absorber pairs with 
galaxy--absorber cross-correlation amplitude $\xi_{\rm ga}(v,\rho) > 1$ and 
with galaxy impact parameter $\rho < 200 \ h^{-1}$ kpc, we confirm and improve 
the results presented by Lanzetta \etal\ (1995) and Chen \etal\ (1998) that (1)
extended gaseous envelopes are a common and generic feature of galaxies of a 
wide range of luminosity and morphological type, (2) the extent of tenuous gas 
($N(\hI) \apg 10^{14}$ \cmjj) around galaxies scales with galaxy $B$-band 
luminosity as $r\propto \,L_B^{0.39\pm 0.09}$, and (3) galaxy interactions do 
not play an important role in distributing tenuous gas around galaxies in most 
cases.  We further demonstrate that (4) the gaseous extent of galaxies scales 
with galaxy $K$-band luminosity as $r\propto \,L_K^{0.28 \pm 0.08}$, and (5) 
tenuous gas around typical $L_*$ galaxies is likely to be distributed in 
spherical halos of radius $\approx 180\ h^{-1}$ kpc of covering factor of 
nearly unity.  The sample consists of 34 galaxy and absorber pairs and 13 
galaxies that do not produce \lya\ absorption lines to within sensitive upper 
limits.  Redshifts of the galaxy and absorber pairs range from $z = 0.0752$ to 
0.8920 with a median of $z = 0.3567$; impact parameter separations of the 
galaxy and absorber pairs range from $\rho = 12.4$ to $175.2 \ h^{-1}$ kpc with
a median of $\rho = 62.2 \ h^{-1}$ kpc.  Of the galaxies, 15 (32\%) are of 
$B$-band luminosity $L_B < 0.25 \ L_{B_*}$ and six (13\%) are of low surface 
brightness.  The galaxy sample is therefore representative of the galaxy 
population over a large fraction of the Hubble time.  Because galaxies of all 
morphological types possess extended gaseous halos and because the extent of 
tenuous gas around galaxies scales with galaxy $K$-band luminosity, we argue 
that galaxy mass---rather than recent star-formation activity---is likely to be
the dominant factor that determines the extent of tenuous gas around galaxies. 
Nevertheless, applying the scaling relationship between the extent of \lya\ 
absorbing gas around galaxies and galaxy $B$-band luminosity, the results of 
our analysis also suggest that the number density evolution of \lya\ absorption
systems may serve to constrain the evolution of the comoving galaxy $B$-band 
luminosity density (at least for the redshift interval between $z\sim 0$ and 
$z\sim 1$ that has been studied in our survey).

\end{abstract}


\newpage

\section{INTRODUCTION}

  The ``forest'' of \lya\ absorption systems observed in the spectra of
background QSOs traces neutral hydrogen gas to redshifts $z\approx 5$.  Because
galaxies in the local universe possess extended neutral hydrogen gas (e.g.\ van
Gorkom 1993), some fraction of the observed \lya\ absorption systems must arise
in individual galaxies.  Whether or not this is a dominant fraction is crucial 
for understanding the origin of \lya\ absorption systems, and understanding the
origin of \lya\ absorption systems bears significantly on all efforts to apply 
the \lya\ forest as a probe of tenuous gas around galaxies over the redshift 
interval probed by the \lya\ absorption systems.  Over the past decade, 
comparison of galaxies and \lya\ absorption systems along common lines of sight
has shown that low-redshift \lya\ absorption systems are associated with 
intervening galaxies (e.g.\ Morris \etal\ 1993), but whether these absorbers 
arise in individual galaxies or merely trace the large-scale galaxy 
distribution is still a matter of some debate (Lanzetta et al.\ 1995; Stocke et
al.\ 1995; Bowen, Blades, \& Pettini 1996; Le Brun, Bergeron, \& Boiss\'e 1996;
van Gorkom et al.\ 1996; Shull, Stocke, \& Penton 1996; Tripp, Lu, \& Savage 
1998; Chen \etal\ 1998; Impey, Petry, \& Flint 1999).  

  Over the past several years, we have been conducting an imaging and 
spectroscopic survey of faint galaxies in fields of Hubble Space Telescope 
(HST) spectroscopic target QSOs (Lanzetta \etal\ 1995; Lanzetta, Webb, \& 
Barcons 1995, 1996, 1997, 2001; Barcons, Lanzetta, \& Webb 1995; Chen \etal\ 
1998).  The goal of the survey is to determine the gaseous extent of galaxies 
and the origin of QSO absorption systems by directly comparing galaxies and 
QSO absorption systems along common lines of sight.  We have so far identified 
352 galaxies of apparent magnitude $m_R < 23$ at redshifts $z < 1.2$, and 230 
\lya\ absorption systems with rest-frame absorption equivalent width $W>0.09$ 
and 36 \civ\ absorption systems with rest-frame absorption equivalent width $W
> 0.09$ at redshifts $z<1.6$ in 24 QSO fields.  Impact parameters (i.e.\ 
projected distances) of the galaxies to the QSO lines of sight range from $\rho
= 10.9 \ h^{-1}$ kpc to $1576.7 \ h^{-1}$ kpc.  The galaxy and absorber pair 
sample is therefore unique for studying the relationship between both \lya\ and
\civ\ absorption systems and galaxies.  In this paper, we address the 
relationship between \lya\ absorption systems and galaxies; we address the
relationship between \civ\ absorption systems and galaxies in a separate paper
(Chen, Lanzetta, \& Webb 2000). 

  The first results of the survey presented by Lanzetta \etal\ (1995) based on 
11 galaxy and \lya\ absorber pairs and 13 galaxies that do not produce 
corresponding \lya\ absorption lines to within sensitive upper limits in six 
QSO fields showed that there is a distinct anti-correlation between \lya\ 
absorption equivalent width and galaxy impact parameter, although the scatter 
about the mean relationship is substantial.  This distinct anti-correlation
strongly indicates that most luminous galaxies possess extended gaseous 
envelopes, but the large scatter about the mean relationship indicates that the
gaseous extent of galaxies may depend on other galaxy properties in addition to
galaxy impact parameter.  To determine how the extent of tenuous gas around 
galaxies scales with galaxy properties is crucial not only for discriminating 
among competing models of the origin of tenuous gas (e.g.\ Rauch 1998 and 
references therein) but also for applying the statistics of \lya\ absorption 
systems to constrain the statistics of faint galaxies over the redshift 
interval probed by the \lya\ absorption systems.

  To address this issue, we have conducted a program to obtain and analyze
images of galaxies in 19 QSO fields, using the HST Wide Field and Planetary
Camera 2 (WFPC2) with the F702W and F606W filters and the NASA 3~m Infrared
Telescope Facility (IRTF) on Mauna Kea with the $K'$ filter.  The primary
objectives of the program are (1) to study how the incidence and extent of
tenuous gas around galaxies depends on galaxy properties, including perhaps
galaxy luminosity, size, or morphological type, (2) to study the spatial
distribution of extended gas around galaxies, e.g.\ whether tenuous gas is
distributed around galaxies in flattened disks (in which case absorption should
occur preferentially in galaxies of low inclination angle) or in spherical
halos (in which case absorption should be independent of galaxy inclination and
orientation), and (3) to determine whether extended gas around galaxies arises 
as a result of galaxy interactions, as evidenced by disturbed morphologies or 
the presence of close companions.  

  Initial results of the program based on 87 galaxies identified in ten QSO 
fields have been published by Chen \etal\ (1998; hereafter Paper I).  In that 
paper, we presented a two-dimensional surface brightness profile analysis of 
HST WFPC2 images of the galaxies to measure galaxy $B$-band luminosity, 
effective radius, mean surface brightness, inclination and orientation of the 
disk component, axial ratio of the bulge component, and disk-to-bulge ratio.  
In addition, we presented an anti-correlation analysis to study the dependence 
of gaseous extent of galaxies on the measurable galaxy parameters.  We found 
that (1) the amount of gas encountered along the line of sight depends on the 
galaxy impact parameter and $B$-band luminosity but does not depend strongly on
the galaxy mean surface brightness, disk-to-bulge ratio, or redshift, (2) 
spherical halos cannot be distinguished from flattened disks based on the 
galaxy and absorber sample, and (3) there is no evidence that galaxy 
interactions play an important role in distributing tenuous gas around galaxies
in most cases.  The statistically significant scaling relation between \lya\ 
absorption equivalent width and galaxy luminosity further supported the
hypothesis that the absorbers trace tenuous gas in individual halos surrounding
the galaxies rather than tenuous gas in galaxy groups or large-scale filaments 
around the galaxies.  

  In this paper, we present complete results of the program, including HST 
WFPC2 photometry for an additional 68 galaxies in the remaining nine QSO fields
and near-infrared photometry for 75 galaxies in 15 of the 19 QSO fields.  The 
new near-infrared galaxy photometry complements the optical galaxy photometry 
and helps to interpret the results of the anti-correlation analysis, because, 
while galaxy $B$-band luminosity is a measure of recent star-forming activity, 
galaxy $K$-band luminosity is a more sensitive measure of total stellar mass,
which may be a more fundamental factor (in comparison with recent star 
formation activity) in scaling the extent of tenuous gas around galaxies.  By
supplementing the optical photometric measurements of the additional galaxies 
and including near-infrared galaxy photometry, we further examine whether or 
not the results of previous analysis may be retained and improved.

  Comparison of galaxies and absorbers identified in the 19 QSO fields yields 
40 galaxies that are associated with corresponding \lya\ absorption lines and 
47 galaxies that do not produce corresponding \lya\ absorption lines to within 
sensitive upper limits.  Galaxy and absorber pairs are considered to be 
physically correlated if (1) the galaxy--absorber cross-correlation amplitude 
satisfies $\xi_{\rm ga}(v,\rho) > 1$ and (2) the galaxy impact parameter 
satisfies $\rho < 200 \ h^{-1}$ kpc.  Including only galaxy and absorber pairs
that are likely to be physically associated and excluding galaxy and absorber
pairs within 3000 \kms\ of the background QSOs leaves 34 galaxy and absorber
pairs and 13 galaxies that do not produce corresponding \lya\ absorption lines
to within sensitive upper limits.  Redshifts of the galaxy and absorber pairs
range from $z = 0.0752$ to 0.8920 with a median of $z = 0.3567$, and impact 
parameter separations of the galaxy and absorber pairs range from $\rho = 12.4$
to $175.2 \ h^{-1}$ kpc with a median of $\rho = 62.2 \ h^{-1}$ kpc.  Of the 47
galaxies, 15 (32\%) are of $B$-band luminosity $L_B < 0.25 \ L_{B_*}$ with 
redshifts ranging from $z = 0.0915$ to 0.6350, and six (13\%) are of low 
surface brightness with redshifts ranging from $z = 0.0915$ to 0.3180.  The 
galaxy sample is therefore representative of the galaxy population over a large
fraction of the Hubble time.

  Based on the new larger galaxy and absorber pair sample, we confirm that the 
amount of gas encountered along the line of sight depends on the galaxy impact 
parameter and $B$-band luminosity but does not depend strongly on the galaxy 
mean surface brightness, disk-to-bulge ratio, or redshift.  In addition, we 
find that (1) the gaseous extent of galaxies scales with galaxy $K$-band 
luminosity as $r\propto\,L_K^{0.28\pm 0.08}$, (2) tenuous gas is more likely to
be distributed in spherical halos than in flattened disks, and (3) typical 
$L_*$ galaxies are surrounded by extended gas of radius $\approx 180\ h^{-1}$ 
kpc and covering factor (within $180 \ h^{-1}$ kpc) of $\approx$ 94\%.  Because
galaxies of all morphological types possess extended gaseous halos and because
the extent of tenuous gas around galaxies scales with galaxy $K$-band 
luminosity, we conclude that galaxy mass---rather than recent star-formation
activity---is likely to be the dominant factor that determines the extent of 
tenuous gas around galaxies.  But because galaxy $K$-band luminosity is
strongly correlated with galaxy $B$-band luminosity (at least for galaxies at
redshifts $z < 1$), we also demonstrate on the basis of the scaling relation 
between the extent of \lya\ absorbing gas around galaxies and galaxy $B$-band 
luminosity that the number density evolution of \lya\ absorption systems may 
serve to constrain the measurements of the cosmic star formation rate density.
We adopt a standard Friedmann cosmology of dimensionless Hubble constant $h = 
H_0/(100$ km s$^{-1}$ Mpc$^{-1})$ and deceleration parameter $q_0 = 0.5$ 
throughout this paper.

\section{OBSERVATIONS}

  In this section, we describe the additional HST WFPC2 observations of 
galaxies in ten QSO fields that were not included in Paper I and near-infrared 
imaging observations of galaxies in 15 QSO fields.

\subsection{WFPC2 Imaging Observations}

  Imaging observations of the fields surrounding 0122$-$0021, 0405$-$1219 
(covering different pointings), 0903$+$1658, 1136$-$1334, 1216$+$0657, 
1259$+$5920, 1424$-$1150, 1641$+$3954, and 2251$+$1552 were obtained with HST 
using WFPC2 with the F702W filter in Cycle 6.  The observations were obtained 
in a series of three exposures of 700 s each.  The journal of observations is 
given in Table 1, which lists the field, 2000 coordinates $\alpha$ and $\delta$
of the QSO, emission redshift $z_{\rm em}$ of the QSO, filter, exposure time, 
and date of observation.

  Imaging observations of objects in the fields surrounding 1317$+$2743 were 
accessed from the HST archive.  The observations were obtained with the HST 
using WFPC2 with the F702W filter.  The observations were carried out in a 
series of four exposures of between 1000 and 1300 s each.  The journal of 
archival observations is given in Table 2, which lists the field, 2000 
coordinates $\alpha$ and $\delta$ of the QSO, emission redshift $z_{\rm em}$ of
the QSO, filter, exposure time, and date of observation.

  All the images were processed following the prescriptions described in Paper 
I.  The spatial resolution of the final images was measured to be ${\rm FWHM} 
\approx 0.1$ arcsec, and the $5 \sigma$ point-source detection thresholds of 
unresolved objects were measured to span the range $m = 26.2$ through $m = 
27.0$. 

\subsection{Near-infrared Imaging Observations}

  Near-infrared imaging observations of objects in the fields 0122$-$0021, 
0349$-$1438, 0405$-$ 1219, 0454$-$2203, 0850$+$4400, 0903$+$1658, 1001$+$2910, 
1136$-$1334, 1216$+$0657, 1259$+$5920, 1354$+$1933, 1424$-$1150, 1545$+$2101,
1704$+$6048, and 2251$+$1552 were obtained using the IRTF 3~m telescope with 
the NSFCAM and the $K'$ filter in April and October 1998. The observations were
carried out in a series of nine exposures dithered by between $\approx 7$ and 
20 arcsec in space to remove hot pixels.  Individual exposures were flat 
fielded using a ``sliding flat'' determined from the median image of nine 
adjacent frames, registered to a common origin using stars or the QSOs, and 
coadded using a proper weight determined from the sky variance to form final 
combined images.  A 1\,$\sigma$ error image was formed simultaneously for each 
combined image through appropriate error propagations.  The total exposure time
of each pointing was approximately 1620 s.  The spatial resolutions of the 
final combined images were measured to span from ${\rm FWHM} \approx 0.6$ 
to $\approx 1.0$ arcsec, and the $5 \sigma$ point-source detection thresholds
were measured to vary from $m = 21.5$ to $m = 22.2$.  

\subsection{Other Observations}

  As described in Paper I, the galaxy and absorber sample is compiled from our
own observations and from observations obtained from the literature.  We 
summarize these observations in Table 3, which for each field lists the number 
of galaxies with spectroscopic redshifts available included into the analysis,
the reference to the galaxy observations and analysis, the number of absorbers
included into the analysis, and the reference to the absorber observations and
analysis.

\section{GALAXY IMAGE ANALYSIS}

  We analyzed all the galaxy images obtained with HST WFPC2 and determined
various galaxy parameters following the procedures described in Paper I.  We 
were able to determine a best-fit surface brightness profile for 142 galaxies, 
but not for 14 galaxies with disturbed morphologies, to which normal disk and 
bulge profiles cannot be applied.  The results of the analysis are disk and 
bulge effective radii, disk-to-bulge ratio, orientation angle, disk inclination
angle, bulge axis ratio, apparent magnitude and mean surface brightness (at the
wavelength centroid of the filter response function) of the galaxies, 
rest-frame $B$-band absolute magnitude $M_B - 5 \log h$, morphological type 
(based on the diagnostics using the disk-to-bulge ratio), and rest-frame 
$B$-band mean surface brightness.

  To determine the near-infrared luminosities of the galaxies, we applied
standard galaxy photometry techniques.  The apparent magnitude $m_K$ was 
measured from the sum of the light within an isophot determined by the 
SExtractor program (Bertin \& Arnouts 1996).  Galaxy fluxes at near-infrared 
wavelengths were calibrated to the standard stars observed every night.  A  
photometric solution was determined for each night using a linear function that
consists of a zero point offset and an extinction coefficient as the two free 
parameters.  The Vega magnitudes of the standard stars in $K$ were converted to
the $AB$ magnitudes according to $AB(K) = K + 1.86$.  Errors of the apparent 
magnitude measurements were determined from the corresponding 1\,$\sigma$ error
images by forming a quadratic sum of pixel values within the isophots.  The 
rest-frame $K$-band absolute magnitude $M_K - 5 \log h$ was determined from the
apparent magnitude $m_K$, corrected for the luminosity distance and the $k$ 
correction.  The rest-frame $K$-band luminosity of an $L_*$ galaxy was taken to
be $M_{K_*} = -21.6$ (Cowie et al.\ 1996). 

  To summarize, we present in Table 4 complete results of measurements for 157 
galaxies in the 19 QSO fields.  In columns (2)---(14) of Table 4, we list for 
each galaxy the field, Right Ascension and Declination offsets from the QSO 
$\Delta \alpha$ and $\Delta \delta$, redshift $z_{\rm gal}$, impact parameter 
$\rho$, disk and bulge effective radii $R_D$ and $R_B$, disk-to-bulge ratio 
$D/B$, orientation angle $\alpha$, disk inclination angle $i$, bulge axis ratio
$b/a$, apparent magnitude $m_{\rm WFPC2}$ (at the wavelength centroid of the 
filter response function), rest-frame $B$-band mean surface brightness $\langle
\mu \rangle$, and absolute $B$-band magnitude $M_B - 5 \log h$.  Measurement 
uncertainties in $R_D$ and $R_B$ were typically 2\%, measurement uncertainties 
in $D/B$ were typically 35\%, measurement uncertainties in $\alpha$ and $i$ 
were typically $2$ deg, and measurement uncertainties in $m_{\rm WFPC2}$ and 
$M_B - 5 \log h$ were typically $0.2$.  In columns(15)---(16) of Table 4, we 
list respectively the apparent and absolute $K$-band magnitudes for 75 
galaxies. 

\section{GALAXY AND ABSORBER PAIRS}

  The goal of the analysis is to investigate tenuous gas around galaxies.  To
accomplish the goal, it is necessary first to distinguish physical pairs from 
correlated and random pairs, which are formed either due to large-scale 
correlation between cluster galaxies and the galaxy that produces the absorber 
or by chance coincidence.  We identify physical galaxy and absorber pairs 
according to the prescription described in Paper I.  First, we accept 
absorption lines according to a $3 \sigma$ detection threshold criterion, which
is appropriate because the measurements are performed at a small number of 
known galaxy redshifts.  Next, we adopt the cross-correlation function 
$\xi_{\rm ga}(v,\rho)$ measured by Lanzetta, Webb, \& Barcons (1997) and form 
galaxy and absorber pairs by requiring (1) $\xi_{\rm ga} > 1$ (which excludes 
likely random pairs) and (2) $\rho < 200 \ h^{-1}$ kpc (which from results of 
Lanzetta, Webb, \& Barcons 1997 excludes likely correlated pairs).  Next, we 
exclude galaxy and absorber pairs within 3000 \kms\ of the background QSOs 
(which are likely to be associated with the QSOs), and in three cases where 
more than one galaxy is paired with one absorber (absorber at $z_{\rm abs} = 
0.3786$ toward 0122$-$0021, absorber at $z_{\rm abs}=0.1670$ toward 
0405$-$1219, and absorber at $z_{\rm abs}=0.4825$ toward 0454$-$2203) we choose
the galaxy at the smallest impact parameter.  Finally, we measure $3 \sigma$ 
upper limits to absorption equivalent widths of galaxies that are not paired 
with corresponding absorbers, retaining only those measurements with $3 \sigma$
upper limits satisfying $W < 0.35$ \AA.  

  This procedure identifies 34 galaxy and \lya\ absorber pairs and 13 galaxies 
that do not produce corresponding \lya\ absorption lines to within sensitive 
upper limits.  Redshifts of the galaxy and \lya\ absorber pairs range from 
0.0752 to 0.8920 with a median of 0.3567, and impact parameter separations of 
the galaxy and absorber pairs range from 12.4 to $175.2 \ h^{-1}$ kpc with a 
median of $62.2 \ h^{-1}$ kpc.  The results are summarized in columns 
(17)---(18) of Table 4, which for each galaxy lists the absorber redshift 
$z_{\rm abs}$, and the rest-frame equivalent width of the \lya\ absorption 
line.  Measurement uncertainties in $W$ are typically 0.1 \AA.  For 
completeness purpose, we also list in column (19) of Table 4 the results of the
same analysis for the \civ\ absorption systems.  The rest-frame \civ\ 
absorption equivalent width is measured for the line at $\lambda_{\rm rest} = 
1548$ \AA.

  In Table 4, galaxy entries without corresponding absorber entries represent 
cases for which the absorption measurement cannot be made, either because the 
galaxy occurs behind the QSO, the appropriate QSO spectrum is not available or 
lacks sensitivity, the spectral region containing the predicted \lya\ or \civ\ 
line is blended with other absorption lines, or a corresponding \lya\ or \civ\ 
absorption line was paired with a galaxy at a smaller impact parameter. 

\section{DESCRIPTIONS OF INDIVIDUAL FIELDS}

  Here we present brief descriptions of galaxies obtained from the new HST 
WFPC2 observations.  We note galaxies by their coordinate offsets in Right 
Ascension and Declination, respectively, from the QSO line of sight in units of
0.1 arcsec.  Individual galaxy images are shown in Figure 1.  The spatial 
extent of each image is roughly $25 \ h^{-1}$ kpc on a side, and orientation of
each image is arbitrary.  Measurements for galaxies published previously in 
Paper I are listed in Table 4 together with the new ones.

\subsection{The Field toward 0122$-$0021}

  Galaxy $-00087$$-00123$ at $z = 0.3788$ and $\rho = 47.1 \ h^{-1}$ kpc 
is a late-type spiral galaxy of luminosity $L_B = 0.69 L_{B_*}$ and color 
$M_B-M_K = 1.3$.  This galaxy is associated with a corresponding \lya\ 
absorption line with $W = 0.74$ \AA\ and a corresponding C\,IV 
absorption line with $W = 0.59$ \AA\ at $z = 0.3786$. The redshift 
determination of this galaxy is uncertain (Q$=$B in Lanzetta \etal\ 1995) and 
is therefore excluded from all analysis.

  Galaxy $-00168$$+00248$ at $z = 0.3992$ and $\rho = 96.2 \ h^{-1}$ kpc is a
late-type spiral galaxy of luminosity $L_B = 1.91 L_{B_*}$ and color 
$M_B-M_K = 1.8$. This galaxy is associated with a corresponding \lya\ 
absorption line with $W = 2.38$ \AA\ and a corresponding C\,IV 
absorption line with $W = 1.70$ \AA\ at $z = 0.3989$.

  Galaxy $+00407$$-00092$ at $z = 0.4299$ and $\rho = 138.9 \ h^{-1}$ kpc is a
late-type spiral galaxy of luminosity $L_B = 0.30 L_{B_*}$ and color 
$M_B-M_K = 1.2$. This galaxy is associated with a corresponding \lya\ 
absorption line at $z = 0.4302$ with $W = 0.84$ \AA\ and does not produce 
corresponding C\,IV absorption to within a sensitive upper limit. The redshift
determination of this galaxy is uncertain (Q$=$B in Lanzetta \etal\ 1995) and 
is therefore excluded from all analysis.

  Galaxy $+00270$$-00372$ at $z = 0.3793$ and $\rho = 143.9 \ h^{-1}$ kpc is a
late-type spiral galaxy of luminosity $L_B = 0.76 L_{B_*}$ and color 
$M_B-M_K = 1.3$.  This galaxy does not have a sensitive \lya\ or C\,IV 
absorption measurement available.

\subsection{The Field toward 0405$-$1219}

  We have published HST WFPC2 images of 14 galaxies surrounding the QSO in 
Paper I.  Here we descriptions for two additional galaxies obtained from the
new observations.

  Galaxy $+00073$$+00036$ at $z = 0.5709$ and $\rho = 30.4 \ h^{-1}$ kpc is a
late-type spiral galaxy of luminosity $L_B = 1.00 L_{B_*}$ and color 
$M_B-M_K = 1.3$.  This galaxy (which occurs in the immediate vicinity of the 
QSO) does not produce corresponding \lya\ or \civ\ absorption to within 
sensitive upper limits.

  Galaxy $-00014$$+00339$ at $z = 0.1670$ and $\rho = 62.8 \ h^{-1}$ kpc shows 
a disturbed morphology, to which normal disk and bulge profiles cannot be 
applied.  Assuming an irregular type galaxy spectral template, we estimate the
$K$-band luminosity of the galaxy to be $L_K = 0.02 L_{K_*}$.  This galaxy is 
associated with a corresponding \lya\ absorption line with $W = 0.65$ \AA\ and 
a corresponding \civ\ absorption line with $W = 0.44$ \AA\ at $z = 0.1670$.  
This absorption system was studied previously by Spinrad \etal\ (1993).  These 
authors attributed this absorption system to a bright, early-type spiral galaxy
$+00405$$-00010$ at $z = 0.1670$ and $\rho = 74.9 \ h^{-1}$ kpc, which fell out
of the WFPC2 frame.  We estimate the $K$-band luminosity of the galaxy to be 
$L_K = 1.20 L_{K_*}$.  It is likely that both galaxies contribute to the \lya\ 
and \civ\ absorption lines (Chen \& Prochaska 2000).

\subsection{The Field toward 0903$+$1658}

  Galaxy $+00066$$-00121$ at $z = 0.4106$ and $\rho = 44.9 \ h^{-1}$ kpc is an 
early-type spiral galaxy of luminosity $L_B = 1.32 L_{B_*}$ and color 
$M_B-M_K = 2.2$. Galaxy $+00370$$+00251$ at $z = 0.4100$ and $\rho = 145.5 
\ h^{-1}$ kpc is an elliptical or S0 galaxy of luminosity $L_B = 0.63 L_{B_*}$
and does not have a $K$-band luminosity measurement available. Galaxies
$-00398$$+00338$ at $z = 0.4087$ and $\rho = 169.7 \ h^{-1}$ kpc and 
$-00390$$+00358$ at $z = 0.4094$ and $\rho = 172.2 \ h^{-1}$ kpc show 
signs of violent interaction, to which normal disk and bulge profiles 
cannot be applied. Galaxy $-00379$$-00506$ at $z = 0.4093$ and $\rho = 205.6 
\ h^{-1}$ kpc is a late-type spiral galaxy of luminosity $L_B = 1.00 L_{B_*}$ 
and color $M_B-M_K = 2.2$. Galaxy $-00236$$+00663$ at $z = 0.4115$ and $\rho = 
229.4 \ h^{-1}$ kpc is a late-type spiral galaxy of luminosity $L_B = 0.58 
L_{B_*}$ and color $M_B-M_K = 1.0$.  These galaxies (which occur in the 
immediate vicinity of the QSO) do not produce corresponding \lya\ or \civ\ 
absorption to within sensitive upper limits.

  Galaxy $+00040$$-00171$ at $z = 0.4258$ and $\rho = 58.2 \ h^{-1}$ kpc is a
late-type spiral galaxy of luminosity $L_B = 0.30 L_{B_*}$ and color 
$M_B-M_K = -0.3$. Galaxy $-00032$$-00240$ at $z = 0.5696$ and $\rho = 80.3 
\ h^{-1}$ kpc shows a disturbed morphology, to which normal disk and bulge 
profiles cannot be applied. These two galaxies occur behind the QSO and so are 
excluded from all analysis.

  Galaxies $+00070$$+00350$ at $z = 0.2690$ and $\rho = 91.8 \ h^{-1}$ kpc,
$-00361$$-00122$ at $z = 0.2695$ and $\rho = 98.1 \ h^{-1}$ kpc,
$-00281$$+00513$ at $z = 0.2697$ and $\rho = 150.7 \ h^{-1}$ kpc, and
$-00287$$+00536$ at $z = 0.2702$ and $\rho = 156.8 \ h^{-1}$ kpc show disturbed
morphologies, to which normal disk and bulge profiles cannot be applied.  
Galaxy $-00178$$+00471$ at $z = 0.2682$ and $\rho = 129.3 \ h^{-1}$ kpc is a 
late-type spiral galaxy of luminosity $L_B = 0.63 L_{B_*}$ and color 
$M_B-M_K = 2.2$. These galaxies do not produce corresponding \civ\ absorption 
to within a sensitive upper limit, but do not have sensitive \lya\ absorption 
measurements available.

\subsection{The Field toward 1136$-$1334}

  Galaxy $-00144$$-00095$ at $z = 0.3191$ and $\rho = 49.2 \ h^{-1}$ kpc is
an elliptical or S0 galaxy of luminosity $L_B = 0.58 L_{B_*}$ and color 
$M_B-M_K = 1.6$.  This galaxy does not produce corresponding \lya\ absorption 
to within a sensitive upper limit, but is associated with a corresponding 
\civ\ absorption line at $z = 0.3189$ with $W = 0.22$ \AA.

  Galaxy $+00004$$+00233$ at $z = 0.5550$ and $\rho = 86.3 \ h^{-1}$ kpc shows
a disturbed morphology, to which normal disk and bulge profiles cannot be 
applied. Galaxy $+00442$$+00091$ at $z = 0.5575$ and $\rho = 167.4 \ h^{-1}$ 
kpc is an elliptical or S0 galaxy of luminosity $L_B = 0.33 L_{B_*}$ and does 
not have a $K$-band luminosity measurement available. These two galaxies 
(which occurs in the immediate vicinity of the QSO) do not produce 
corresponding \lya\ or \civ\ absorption to within sensitive upper limits.

  Galaxy $+00108$$-00255$ at $z = 0.2044$ and $\rho = 59.4 \ h^{-1}$ kpc is a
late-type spiral galaxy of luminosities $L_B = 0.33 L_{B_*}$ and color 
$M_B-M_K = 2.3$. This galaxy is associated with a corresponding \civ\ 
absorption line at $z = 0.2039$ with $W = 0.76$ \AA, but does not have a 
sensitive \lya\ absorption measurement available.

  Galaxy $-00271$$+00515$ at $z = 0.3604$ and $\rho = 177.4 \ h^{-1}$ kpc is 
a late-type spiral galaxy of luminosity $L_B = 0.28 L_{B_*}$ and color 
$M_B-M_K = 2.0$. Galaxy $-00073$$-00807$ at $z = 0.3254$ and $\rho = 233.5 
\ h^{-1}$ kpc is a late-type spiral galaxy of luminosity $L_B = 1.45 L_{B_*}$
and does not have a $K$-band luminosity measurement available. Galaxies 
$-00365$$-00012$ at $z = 0.2198$ and $\rho = 82.3 \ h^{-1}$ kpc, 
$-00182$$-00483$ at $z = 0.2123$ and $\rho = 113.6 \ h^{-1}$ kpc, and 
$-00520$$+00138$ at $z = 0.3598$ and $\rho = 163.8 \ h^{-1}$ kpc are 
early-type spiral galaxies of luminosities between $L_B = 0.03 L_{B_*}$ and 
$L_B = 2.29 L_{B_*}$ and colors between $M_B-M_K = 2.1$ and $M_B-M_K = 2.2$. 
Galaxy $-00451$$+00388$ at $z = 0.3595$ and $\rho = 181.1 \ h^{-1}$ kpc is an 
elliptical or S0 galaxy of luminosity $L_B = 1.58 L_{B_*}$ and color 
$M_B-M_K = 2.9$. Galaxy $-00032$$+00557$ at $z = 0.1755$ and $\rho = 107.2 
\ h^{-1}$ kpc shows a disturbed morphology, to which normal disk and bulge 
profiles cannot be applied. These galaxies do not produce corresponding \civ\ 
absorption to within sensitive upper limits, but do not have sensitive \lya\ 
absorption measurements available.

  Galaxy $+00019$$-00371$ at $z = 0.6480$ and $\rho = 144.8 \ h^{-1}$ kpc is
an elliptical or S0 galaxy of luminosity $L_B = 1.45 L_{B_*}$ and color 
$M_B-M_K = 1.8$. This galaxy occurs behind the QSO and so is excluded from all 
analysis.

  Galaxy $+00178$$-00542$ at $z = 0.4007$ and $\rho = 183.6 \ h^{-1}$ kpc is
a late-type spiral galaxy of luminosity $L_B = 0.52 L_{B_*}$ and color 
$M_B-M_K = 0.9$.  This galaxy does not produce corresponding \lya\ or \civ\ 
absorption to within a sensitive upper limit. The redshift determination of 
this galaxy is uncertain (Q$=$B in Lanzetta \etal\ 1995) and is therefore 
excluded from all analysis.

\subsection{The Field toward 1216$+$0657}

  Galaxy $+00372$$-00188$ at $z = 0.1242$ and $\rho = 61.3 \ h^{-1}$ kpc is a 
late-type spiral galaxy of luminosity $L_B = 0.63 L_{B_*}$ and color 
$M_B-M_K = 1.2$.  This galaxy is associated with a corresponding \lya\ 
absorption line with $W = 1.26$ \AA\ and a corresponding \civ\ 
absorption line with $W = 0.31$ \AA at $z = 0.1243$.

  Galaxy $-00169$$-00569$ at $z = 0.6021$ and $\rho = 226.1 \ h^{-1}$ kpc is
a late-type spiral galaxy of luminosity $L_B = 1.58 L_{B_*}$ and color 
$M_B-M_K = 1.4$.  Galaxy $-00186$$-00893$ at $z = 3.2720$ and $\rho = 320.4 
\ h^{-1}$ kpc shows a compact morphology with a brightness profile best 
represented by the $R^{1/4}$ law and it is very bright. Therefore, it is 
likely to be a QSO.  Galaxy $+00496$$-00775$ at $z = 0.4341$ and $\rho = 307.7 
\ h^{-1}$ kpc is an early-type spiral galaxy of luminosity $L_B = 0.58 L_{B_*}$
and does not have a $K$-band luminosity measurement available. These objects 
occur behind the QSO and so are excluded from all analysis.

  Galaxy $-00606$$-00774$ at $z = 0.0012$ and $\rho = 1.7 \ h^{-1}$ kpc is an
elliptical or S0 galaxy of luminosity $L_B \ll\ 0.01 L_{B_*}$ and does not 
have a $K$-band luminosity measurement available. This galaxy does not have a 
sensitive \lya\ or \civ\ absorption measurement available.

\subsection{The Field toward 1259$+$5920}

  Galaxy $+00270$$-00313$ at $z = 0.1967$ and $\rho = 86.2 \ h^{-1}$ kpc is an 
elliptical or S0 galaxy of luminosity $L_B = 0.10 L_{B_*}$ and color 
$M_B-M_K = 1.6$.  This galaxy is associated with a corresponding \lya\ 
absorption line at $z = 0.1966$ with $W = 0.22$ \AA, but does not produce 
corresponding \civ\ absorption to within a sensitive upper limit.

  Galaxy $-00605$$+00039$ at $z = 0.5353$ and $\rho = 221.5 \ h^{-1}$ kpc 
ia a late-type spiral galaxy of luminosity $L_B = 1.00 L_{B_*}$ and does not 
have a $K$-band luminosity measurement available. Galaxy $+00087$$+00110$ at 
$z = 0.4869$ and $\rho = 391.3 \ h^{-1}$ kpc is a late-type spiral galaxy of 
luminosity $L_B = 0.83 L_{B_*}$ and color $M_B-M_K = 1.5$.  These two galaxies 
occur behind the QSO and so are excluded from all analysis.

  Galaxy $-00234$$+00685$ at $z = 0.2412$ and $\rho = 173.6 \ h^{-1}$ kpc is
a late-type spiral galaxy of luminosity $L_B = 0.48 L_{B_*}$ and color 
$M_B-M_K = 1.2$.  This galaxy does not produce corresponding \lya\ or \civ\ 
absorption to within a sensitive upper limit.

\subsection{The Field toward 1317$+$2743}

  Galaxy $+00068$$+00048$ at $z = 0.6715$ and $\rho = 32.8 \ h^{-1}$ kpc is an 
elliptical or S0 galaxy of luminosity $L_B = 1.32 L_{B_*}$ and does not 
have a $K$-band luminosity measurement available.  This galaxy is
associated with a corresponding \lya\ absorption line at $z = 0.6716$ with
$W = 0.78$ \AA, but does not have a sensitive \civ\ absorption measurement 
available.

  Galaxy $-00444$$-00023$ at $z = 0.6717$ and $\rho = 175.2 \ h^{-1}$ kpc is 
an elliptical or S0 galaxy of luminosity $L_B = 1.00 L_{B_*}$ and does not 
have a $K$-band luminosity measurement available.  This galaxy is
associated with a corresponding \lya\ absorption line at $z = 0.6736$ with
$W = 0.78$ \AA, but does not produce corresponding \civ\ absorption to within 
a sensitive upper limit.

  Galaxies $-00661$$+00105$ at $z = 0.5397$ and $\rho = 245.3 \ h^{-1}$ kpc and
$-00681$$+00097$ at $z = 0.5398$ and $\rho = 252.1 \ h^{-1}$ kpc show 
signs of violent interaction, to which normal disk and bulge profiles 
cannot be applied.  These two galaxies do not produce corresponding \lya\ 
absorption to within a sensitive upper limit, but do not have sensitive \civ\ 
absorption measurements available.

\subsection{The Field toward 1424$-$1150}

  Galaxy $+00030$$-00013$ at $z = 0.8011$ and $\rho = 13.4 \ h^{-1}$ kpc is a
late-type spiral galaxy of luminosity $L_B = 1.20 L_{B_*}$ and does not 
have a $K$-band luminosity measurement available. This galaxy (which 
occur in the immediate vicinity of the QSO) does not produce corresponding 
\lya\ or \civ\ absorption to within a sensitive upper limit.

  Galaxy $-00002$$+00176$ at $z = 0.3404$ and $\rho = 52.0 \ h^{-1}$ kpc is a 
late-type spiral galaxy of luminosity $L_B = 0.63 L_{B_*}$ and color 
$M_B-M_K = 2.0$.  This galaxy is associated with a corresponding \lya\ 
absorption line at $z = 0.3417$ with $W = 0.60$ \AA, but does not produce 
corresponding \civ\ absorption to within a sensitive upper limit.

  Galaxy $-00114$$+00409$ at $z = 0.1064$ and $\rho = 55.0 \ h^{-1}$ kpc is a
late-type spiral galaxy of luminosity $L_B = 0.02 L_{B_*}$ and does not 
have a $K$-band luminosity measurement available. Galaxy
$-00178$$+00864$ at $z = 0.1038$ and $\rho = 111.9 \ h^{-1}$ kpc is an 
elliptical or S0 galaxy of luminosity $L_B = 0.13 L_{B_*}$ and color 
$M_B-M_K = 2.3$. Galaxies $-00881$$-00478$ at $z = 0.1045$ and $\rho = 127.9 
\ h^{-1}$ kpc and $-00913$$-00538$ at $z = 0.2597$ and $\rho = 266.6 \ h^{-1}$ 
kpc are early-type spiral galaxies of luminosities $L_B = 0.13 L_{B_*}$ and 
$L_B = 0.36 L_{B_*}$, respectively, and do not have $K$-band luminosity 
measurements available.  These galaxies do not have sensitive \lya\ or \civ\ 
absorption measurements available.

  Galaxy $-00691$$-00339$ at $z = 0.3942$ and $\rho = 245.7 \ h^{-1}$ kpc is
a late-type spiral galaxy of luminosity $L_B = 0.25 L_{B_*}$ and does not 
have a $K$-band luminosity measurement available.  This galaxy 
does not produce corresponding \lya\ or \civ\ absorption to within a sensitive 
upper limit.

\subsection{The Field toward 1641$+$3954}

  Galaxies $+00068$$-00013$ at $z = 0.5880$ and $\rho = 26.2 \ h^{-1}$ kpc, 
$-00108$$+00279$ at $z = 0.5900$ and $\rho = 113.2 \ h^{-1}$ kpc, 
$-00366$$+00366$ at $z = 0.5934$ and $\rho = 196.2 \ h^{-1}$ kpc, and 
$-00621$$+00065$ at $z = 0.5917$ and $\rho = 236.5 \ h^{-1}$ kpc are late-type 
spiral galaxies of luminosities in the range $L_B = 1.00 L_{B_*}$  to 
$1.10 L_{B_*}$.  Galaxies $+00289$$-00369$ at $z = 0.5918$ and 
$\rho = 177.5 \ h^{-1}$ kpc is an elliptical or S0 galaxies of luminosity 
$L_B = 1.45 L_{B_*}$.  These galaxies (which occur in the immediate vicinity 
of the QSO) do not produce corresponding \lya\ or \civ\ absorption to within a 
sensitive upper limit.

  Galaxies $+00122$$-00189$ at $z = 0.2813$ and $\rho = 59.5 \ h^{-1}$ kpc, 
$-00208$$+00107$ at $z = 0.4126$ and $\rho = 76.4 \ h^{-1}$ kpc,
$+00322$$+00625$ at $z = 0.2410$ and $\rho = 168.5 \ h^{-1}$ kpc, and
$+00256$$-00668$ at $z = 0.2192$ and $\rho = 160.9 \ h^{-1}$ kpc
are late-type spiral galaxies of luminosities between $L_B = 0.04 L_{B_*}$ and
$0.36 L_{B_*}$.  Galaxies $+00406$$-00388$ at $z = 0.3316$ and $\rho = 163.6 
\ h^{-1}$ kpc, and $-00506$$-00327$ at $z = 0.4055$ and $\rho = 195.0 
\ h^{-1}$ kpc are elliptical or S0 galaxies of luminosities $L_B = 0.40 
L_{B_*}$. These galaxies do not have sensitive \lya\ or \civ\ absorption 
measurements available.

  Galaxy $-00164$$-00157$ at $z = 0.5319$ and $\rho = 82.7 \ h^{-1}$ kpc is a
late-type spiral galaxy of luminosity $L_B = 0.76 L_{B_*}$.  This galaxy is
associated with a corresponding \lya\  absorption line at $z  = 0.5342$ with 
$W = 1.11$ \AA, but does not have a sensitive \civ\ absorption measurement 
available.

  Galaxy $-00441$$+00323$ at $z = 0.6944$ and $\rho = 217.4 \ h^{-1}$ kpc is
a late-type spiral galaxy of luminosity $L_B = 2.09 L_{B_*}$. This galaxy
occurs behind the QSO and so is excluded from all analysis.

  Galaxies $-00183$$+00646$ at $z = 0.5296$ and $\rho = 244.3 \ h^{-1}$ kpc and
$+00321$$-00834$ at $z = 0.5289$ and $\rho = 325.0 \ h^{-1}$ kpc are elliptical
or S0 galaxies of luminosities $L_B = 0.83 L_{B_*}$ and $L_B = 1.20 L_{B_*}$,
respectively.  These two galaxies do not produce corresponding \lya\ 
absorption to within a sensitive upper limit, but do not have sensitive \civ\ 
absorption measurements available.

  We do not have $K$-band photometry for this field.

\subsection{The Field toward 2251$+$1552}

  Galaxy $+00322$$-00003$ at $z = 0.3529$ and $\rho = 97.0 \ h^{-1}$ kpc is a
late-type spiral galaxy of luminosity $L_B = 0.63 L_{B_*}$ and color 
$M_B-M_K = 1.3$.  This galaxy is associated with a corresponding \lya\ 
absorption line at $z = 0.3526$ with $W = 0.70$ \AA, but does not produce 
corresponding \civ\ absorption to within a sensitive upper limit.

\section{ANALYSIS}

  In order to determine how the extent of tenuous gas around galaxies depends 
on galaxy properties, we performed in Paper I a maximum likelihood analysis to 
(1) confirm the existence of a fiducial relationship between some measure of 
the strength of neutral hydrogen absorption (e.g.\ \lya\ absorption equivalent 
width $W$ or neutral hydrogen column density $N$) and galaxy impact parameter 
and (2) assess whether accounting for measurements of other galaxy properties 
(such as galaxy $B$-band luminosity $L_B$, effective radius $r_e$, mean surface
brightness $\langle \mu \rangle$, disk-to-bulge ratio $D/B$, or redshift $z$) 
can improve upon the fiducial relationship.  Here we examine whether or not the
results of previous analysis can be retained and improved by including 
additional measurements.  First, we consider two different geometries of gas 
distribution---a spherical halo model and a flattened disk model---and repeat 
the maximum-likelihood analysis based on the new larger galaxy and absorber 
pair sample to determine the best-fit model and the intrinsic variation, 
$\sigma_c$.  Next, we determine whether accounting for galaxy near-infrared 
photometry can improve the fiducial relationship between \lya\ absorption 
equivalent width and galaxy impact parameter.  Finally, we continue to examine 
the HST WFPC2 images of the additional galaxies to determine whether extended 
gas arises as a result of galaxy interactions, as evidenced by disturbed 
morphologies or the presence of close companions.

\subsection{The Relation between Galaxies and \lya\ Absorption Systems}

  To confirm the existence of a fiducial relationship between some measure of 
the strength of neutral hydrogen absorption (e.g.\ \lya\ absorption equivalent 
width $W$ or neutral hydrogen column density $N$) and galaxy impact parameter
$\rho$, we adopt the parameterized linear form
\begin{equation}
y = a_1 x_1 + {\rm constant}.
\end{equation}
The ``dependent measurement'' $y$ represents the strength of neutral hydrogen 
absorption and may be $\log W$, $\log\,(W \cos i)$, $\log N$, or $\log\,(N 
\cos i)$.  (The $\cos i$ factor is included as a path-length correction in 
models of inclined galaxy disks.)  The ``independent measurement'' $x_1$
represents some measure of the distance between the absorber and the galaxy
and may be $\log \rho$ for a spherical halo model or $\log R$ for a flattened
disk model.  Here the galactocentric radius $R$ is related to the galaxy impact
parameter $\rho$ by
\begin{equation}
R = \rho \left[ 1 + \sin^2 \alpha \tan^2 i \right]^{1/2},
\end{equation}
where $\alpha$ is the orientation angle between the apparent major axis of the 
galaxy and the projected line segment joining the galaxy to the QSO.

  To assess whether accounting for measurements of other galaxy properties 
(in addition to galaxy impact parameter) can improve upon the fiducial 
relationship, we adopt the parameterized bi-linear form
\begin{equation}
y = a_1 x_1 + a_2 x_2 + {\rm constant}.
\end{equation}
The additional ``independent measurement'' $x_2$ represents galaxy properties 
and may be $\log L_B$, $\log r_e$, $\langle \mu \rangle$, $\log D/B$, or $\log 
(1+z)$.  The goodness of fit for each model is estimated based on a 
``confidence interval test'' and an ``anti-correlation test''.  Results of the 
statistical tests of each model is presented in Table 5, which lists the 
measurements; the statistical significances $a/\delta a$ of the fitting 
coefficients; the correlation coefficients $r_{gk}$, $r_k$, $r_s$, and $r_p$ 
and the corresponding statistical significances $r_{gk}/\sigma_{gk}$, 
$r_k/\sigma_k$, $r_s/\sigma_s$, and $r_p/\sigma_p$ of the generalized Kendall, 
Kendall, Spearman, and Pearson correlation tests, respectively; and the cosmic 
scatter $\sigma_c$.

  Comparison of the results in Table 5 and the ones presented in Paper I shows 
that (1) the anti-correlation between \lya\ absorption equivalent width and 
galaxy impact parameter remains at a higher significance level, (2) including 
galaxy $B$-band luminosity, effective radius, or redshift as an additional 
scaling factor continues to substantially improve the $W$ versus $\rho$ 
anti-correlation, and (3) including galaxy mean surface brightness or 
disk-to-bulge ratio as an additional scaling factor remains statistically 
identical to the fiducial relationship between $W$ and $\rho$.  The similarity 
between the results accounting for galaxy $B$-band luminosity and the ones 
accounting for galaxy effective radius may be attributed to the Holmberg (1975)
relation between galaxy luminosity and size, and the similarity between the 
results accounting for galaxy $B$-band luminosity and the ones accounting for 
galaxy redshift to the selection effect due to a magnitude-limited survey.  The
latter is demonstrated in Figure 1 of Chen, Lanzetta, \& Fern\'andez-Soto 
(2000), which shows the residuals of the $W$ vs. $\rho$ anti-correlation after 
accounting for galaxy $B$-band luminosity as a function of galaxy redshift.  No
correlation is found between the residuals and galaxy redshifts.  Therefore, we
confirm that the amount of gas intercepted along the line of sight depends on 
galaxy impact parameter and $B$-band luminosity, but does not depend strongly 
on galaxy mean surface brightness, disk-to-bulge ratio, or redshift. 

  To illustrate the dependence/independence of \lya\ absorption equivalent 
width on various galaxy parameters, we plot in Figure 1 the residuals of the 
$W$ versus $\rho$ anti-correlation as a function of galaxy $B$-band luminosity 
(the upper-left panel), redshift (the upper-right panel), galaxy mean surface 
brightness (the lower-left panel), and galaxy disk-to-bulge ratio (the 
lower-right panel).  Circles represent elliptical or S0 galaxies; triangles 
represent early-type spiral galaxies; and squares represent late-type spiral 
galaxies.  Points with arrows indicate $3 \sigma$ upper limits.  The residuals 
appear to correlate strongly only with galaxy $B$-band luminosity, but not with
galaxy redshift, mean surface brightness, or disk-to-bulge ratio.  Note that 
Figure 1 also indicates the survey range of the galaxy and absorber pair sample
in various galaxy parameter spaces.  Specifically, of all the 47 galaxies in 
the sample, two are of luminosities $L_B < 0.04 L_{B_*}$ at redshifts $z = 
0.0949$ and $z = 0.1380$, respectively, and 13 are of luminosities $0.04\ 
L_{B_*} \apl L_B \apl 0.25 \ L_{B_*}$ with redshifts spanning from $z = 0.0915$
to $z = 0.6350$.  Namely, 32\% of the sample is made up of faint dwarf galaxies
that span a wide redshift range.  Furthermore, the dotted line in the 
lower-left panel devides between low surface brightness galaxies and high 
surface brightness ones.\footnote{The value is determined by adopting a 
$B$-band central surface brightness, $\mu_0 = 23.0$ mag sec$^{-2}$ and 
transforming to the corresponding mean surface brightness within the Holmberg 
radius, assuming an exponential disk profile.}  Therefore, six of the 47 
galaxies (13\%) are low surface brightness galaxies at redshifts between $z = 
0.0915$ to $z = 0.3180$.  Finally, the lower-right panel indicates that 
galaxies in the sample span a wide range in the disk-to-bulge ratio, from 
bulge-dominated galaxies to disk-dominated galaxies.

  To study whether tenuous gas is distributed around galaxies in flattened 
disks, in which case the absorption signatures should occur preferentially in 
galaxies of low inclination angles, or in spherical halos, in which case the 
absorption signatures should be independent of galaxy inclination and 
orientation, we compare the statistical significances of the $W$ versus $\rho$ 
and $W$ versus $R$ anti-correlations.  In Paper I, we were unable to 
distinguish between a spherical halo model and a flattened disk model because 
of the statistically identical results of the tests.  The results presented in 
Table 5 based on the larger galaxy and absorber sample, however, indicate that 
although the strong anti-correlations between $W$ and $R$ remain (rows 7 to 
13), they are now marginally inferior to the $W$ versus $\rho$ 
anti-correlations (rows 1 to 6).  Specifically, the anti-correlation test 
indicates that the anti-correlation between $W$ and $\rho$ is at a level of 
significance ranging from $4.1\ \sigma$ to $5.3\ \sigma$, while the 
anti-correlations between $W\cos i$ and $R$ is at a level of significance 
ranging from $3.2\ \sigma$ to $4.0\ \sigma$; the anti-correlation between $W$, 
$\rho$, and $L_B$ is at a level of significance ranging from $5.2\ \sigma$ to 
$7.1\ \sigma$, while the anti-correlations between $W\cos i$, $R$, and $L_B$ is
 at a level of significance ranging from $4.7\ \sigma$ to $6.2\ \sigma$.  Given
that several highly inclined disk galaxies with a 90-degree orientation angle 
to the QSOs are directly observed in our sample (e.g.\ galaxy $+00106$$+00057$ 
toward 0454$-$2203 and galaxy $-00042$$-00038$ toward 1622$+$2352), we 
attribute the $W$ versus $R$ anti-correlation to the correlation between $\rho$
and $R$ defined in Equation (2) and conclude that tenuous gas is more likely to
be distributed in a spherical halo than in a flattened disk.  

  To summarize, the results of the likelyhood analysis show that tenuous gas
around galaxies may be described by 
\begin{equation}
\log W = -(0.96\pm 0.11) \log \rho + {\rm constant},
\end{equation}
and is better described by
\begin{equation}
\log W = -\alpha \log \rho + \beta \log L_B + {\rm constant},
\end{equation}
where
\begin{equation}
\alpha = 1.04 \pm 0.10
\end{equation}
and
\begin{equation}
\beta = 0.40 \pm 0.09.
\end{equation}
Roughly relating \lya\ absorption equivalent width $W$ to neutral hydrogen
column density $N$ according to the prescription described by Lanzetta et al.\
(1995) and in Paper I, tenuous gas around galaxies may also be described by 
\begin{equation}
\log \left( \frac{N}{10^{20} \ \cmjj} \right)  = -\alpha \log \left(
\frac{\rho}{10 \ {\rm kpc}} \right) + \beta \log \left( \frac{L_B}{L_{B_*}}
\right) + {\rm constant}
\end{equation}
where
\begin{equation}
\alpha = 5.56 \pm 0.42,
\end{equation}
\begin{equation}
\beta = 2.56 \pm 0.43,
\end{equation}
and
\begin{equation}
{\rm constant} = 1.56 \pm 0.78.
\end{equation}
Comparisons of the data and the best-fit models are presented in Figures 3, 4 
and 5 with the estimated cosmic scatter of each relationship shown in the 
upper-right corner.

\subsection{Gaseous Extent as a Function of Galaxy $K$-Band Luminosity}

  We have confirmed in \S\S\ 6.1 that the extent of tenuous gas around galaxies
scales with galaxy $B$-band luminosity.  Interpreting this scaling relation is 
difficult, because $B$-band luminosity is sensitive to recent star formation 
activity, which is expected to be stronger in galaxies of later types, but our
analysis also indicates that the extent of tenuous gas around galaxies is 
insensitive to galaxy morphology.  It appears that galaxy $B$-band luminosity 
may not be a true fundamental parameter.  Because $K$-band luminosity is 
sensitive to the old stellar content and is considered to be a good measure of 
the total stellar mass (Bruzual \& Charlot 1993), here we consider the 
possibility that the amount of gas intercepted along the line of sight depends 
on galaxy rest-frame $K$-band luminosity $L_K$.

  Considering only galaxies with $K$-band photometric measurements available
leaves 20 galaxy and absorber pairs and two galaxies that do not produce 
corresponding \lya\ absorption lines to within sensitive upper limits.  
Adopting a power-law relationship between $W$ and $\rho$ and $L_K$, we find
according to the likelihood analysis that tenuous gas around galaxies may be
described by
\begin{equation}
\log W = -\alpha \log \rho + \beta_k \log L_K + {\rm constant}
\end{equation}
where
\begin{equation}
\alpha = 1.25 \pm 0.18
\end{equation}
and
\begin{equation}
\beta_k = 0.35 \pm 0.08.
\end{equation}
The result applies over the $K$-band luminosity interval $0.005\ L_{K_*} \apl 
L_K \apl 3.3\ L_{K_*}$ spanned by the observations.  Comparison of the data and
the best-fit model is shown in the left panel of Figure 6.  

  For comparison, we repeat the likelihood analysis for the subsample, but
replacing galaxy $K$-band luminosity with galaxy $B$-band luminosity.  Results
of the analysis yield $\alpha = 1.16 \pm 0.17$ and $\beta = 0.50 \pm 0.11$.  
The result applies over the $B$-band luminosity interval $0.04\ L_{B_*} \apl 
L_B \apl 2.6\ L_{B_*}$ spanned by the subsample.  Comparison of the data and 
the best-fit model is shown in the right panel of Figure 6.  The results of the
statistical tests for the subsample are presented in Table 6.  

  We find that the $W$ versus $\rho$ anti-correlation accounting for galaxy 
$K$-band luminosity and the one accounting for galaxy $B$-band luminosity are 
both superior to the fiducial $W$ versus $\rho$ anti-correlation, indicating 
that the extent of tenuous gas around galaxies depends sensitively on galaxy 
$K$-band luminosity.  In addition, although the results of the statistical 
tests indicate that the $W$ versus $\rho$ anti-correlation accounting for $L_B$
is statistically comparable to the $W$ versus $\rho$ anti-correlation 
accounting for $L_K$, the cosmic scatter $\sigma_c$ in the latter case is 
further reduced by a substantial amount (33\%).  Because of the marginally 
stronger scaling relation between the gaseous extent of galaxies and galaxy 
$K$-band luminosity and because galaxies of all morphological types possess 
extended gaseous halos, we find that galaxy mass is likely to be the dominant 
factor that determines the extent of tenuous gas around galaxies.  The 
similarity between the results accounting for galaxy $B$-band luminosity and 
the ones accounting for galaxy $K$-band luminosity may be attributed to the 
strong correlation between galaxy $K$-band and $B$-band luminosities for 
galaxies in the sample as shown in Figure 7.

\subsection{Galaxy Interactions}

  By carefully examining HST WFPC2 images of galaxies in our sample, we found 
in Paper I no evidence that galaxy interactions play an important role in 
distributing tenuous gas around galaxies in all cases.  Here we investigate  
whether this result remains valid for the additional galaxies in the new nine 
QSO fields.  We examine whether or not galaxies in the new sample exhibit close
companions or disturbed morphologies in the HST WFPC2 images.  Of all the 
galaxies presented in Figure 1, only galaxy $-00014$$+00339$ toward 0405$-$1219
appears to exhibit obvious signs of a disturbed morphology and is associated 
with a corresponding \lya\ absorption line at $z=0.167$.  (But galaxy 
$+00405$$-00010$ toward 0405$-$1219 also occurs at the same redshift and is
likely to contribute to the \lya\ absorption system.)  Galaxy $-00032$$+00557$ 
toward 1136$-$1334 and galaxies $+00070$$+00350$, $-00361$$-00122$, 
$-00281$$+00513$, and $-00287$$+00536$ toward 0903$+$1658, which occur at small
impact parameters (between $\rho = 80.3 \ h^{-1}$ kpc and $\rho = 156.8 \ 
h^{-1}$ kpc), appear to have disturbed morphologies but do not have sensitive 
\lya\ absorption measurements available.  Galaxy $+00004$$+00233$ toward 
1136$-$1334, and galaxies $-00398$$+00338$ and $-00390$$+00358$ toward 
0903$+$1658, which occur in the immediate vicinity of the QSOs, appear to have 
disturbed morphologies and do not produce corresponding \lya\ absorption lines 
to within a sensitive upper limit.  The galaxy pair $-00661$$+00105$ and 
$-00681$$+00097$ at $z\approx 0.5398$ toward 1317$+$2743, which occurs at 
relatively large impact parameters ($\rho \approx\ 250 \ h^{-1}$ kpc), appears 
to have disturbed morphologies and does not produce a corresponding \lya\ 
absorption line to within a sensitive upper limit.  Based on the new galaxy 
sample, we confirm that there is no evidence that tenuous gas is distributed 
around galaxies as a result of galaxy interactions in most cases, although we 
cannot rule out the possibility that tenuous gas is distributed around galaxies
as a result of galaxy interactions in some cases.

  Additional support may be inferred based on the results shown in the 
lower-right panel of Figure 2 that the residuals of the $W$ vs. $\rho$ 
relationship do not correlate strongly with galaxy morphology as represented by
the disk-to-bulge ratio.  Given the apparent morphology-density relation 
observed for galaxies in the local universe (e.g. Dressler et al. 1997), we 
find it unlikely that gaseous extent of galaxies is correlated with the 
surrounding galaxy environment.

\section{DISCUSSION}

  Adopting a quantitative criterion to identify physical galaxy and absorber 
pairs, we have repeated the likelihood analysis to study the relation between 
\lya\ absorption systems and galaxies.  Considering only galaxy and absorber 
pairs that are likely to be physically associated and excluding galaxy and 
absorber pairs within 3000 \kms\ of the background QSOs leaves 34 galaxy and 
absorber pairs and 13 galaxies that do not produce corresponding \lya\ 
absorption lines to within sensitive upper limits.  We confirm that the amount 
of gas encountered along the line of sight depends on the galaxy impact 
parameter $\rho$ and galaxy $B$-band luminosity $L_B$ but does not depend 
strongly on the galaxy mean surface brightness $\langle \mu \rangle$, 
disk-to-bulge ratio $D/B$, or redshift $z$ and that there is no evidence that 
galaxy interactions play an important role in distributing tenuous gas around 
galaxies in most cases.  We also demonstrate that the amount of gas encountered
along the line of sight also depends on galaxy $K$-band luminosity $L_K$ and 
that tenuous gas is likely to be distributed in spherical halos, rather than 
in flattened disks.

  The statistically significant anti-correlation between \lya\ absorption 
equivalent width and galaxy impact parameter strongly supports that most 
absorbers are indeed associated with the individual galaxies identified.  Most 
importantly, we show that extended gaseous halos are a common and generic 
feature of galaxies of all morphological types and that galaxy mass is the 
dominant factor that determines the extent of tenuous gas around galaxies.  The
strong scaling relationship between \lya\ absorption equivalent width and 
galaxy luminosity further supports that the absorbers trace tenuous gas in 
individual halos surrounding the galaxies rather than tenuous gas in galaxy 
groups or large-scale filaments around the galaxies.  Although we cannot 
exclude the possibility that the absorbers may arise in a population of dwarfs
bound in a larger potential well, the scaling relationship strongly argues that
the absorption gas cross section is determined by the luminous galaxies 
identified in our survey.  In addition, Bothun et al. (1993) have shown that 
low surface brightness disk galaxies (with scale size between 1 and 5 kpc) tend
to avoid virialized regions and are less clustered.  It would be therefore very
uncharacteristic to find low surface brightness galaxies in addition to the 
galaxies identified in our survey within a radius of 200 kpc from the sightline
and $\pm 250$ \kms away from the absorber redshifts.  Here we discuss the
implications drawn from the results of our analysis.

\subsection{The Incidence and Extent of Tenuous Gas Around Galaxies}

  Adopting the results of the likelihood analysis presented in \S\S\ 6.1, 
we update the scaling relationship between the gaseous extent $r$ of galaxies 
and galaxy $B$-band luminosity $L_B$.  We find in a complete agreement with our
previous analysis that the extent of tenuous gas around galaxies scales with 
galaxy $B$-band luminosity by
\begin{equation}
\frac{r}{r_*} = \left( \frac{L_B}{L_{B_*}} \right)^{t_B},
\end{equation}
where $t_B = \beta / \alpha$ and is estimated based on Equations (6) and (7) to
be
\begin{equation}
t_B = 0.39 \pm 0.09,
\end{equation}
and that the gaseous extent of a typical $L_*$ galaxy is 
\begin{equation}
r_* = 184_{-25}^{+29} \ h^{-1} \ {\rm kpc}
\end{equation}
at a \lya\ absorption equivalent width threshold $W = 0.3$ \AA.  
The results apply over the $B$-band luminosity interval $0.03 \apl L_B \apl 
2.6 L_{B_*}$ and for galaxies of all morphological types spanned by the 
observations.

  Adopting the results of the likelihood analysis presented in \S\S\ 6.2, we
also determine the scaling relationship between the gaseous extent of galaxies 
and galaxy $K$-band luminosity $L_K$.  We find that the extent of tenuous gas 
around galaxies scales with galaxy $K$-band luminosity by
\begin{equation}
\frac{r}{r_*} = \left( \frac{L_K}{L_{K_*}} \right)^{t_K},
\end{equation}
where 
\begin{equation}
t_K = 0.28 \pm 0.08
\end{equation}
and
\begin{equation}
r_* = 177_{-26}^{+30} \ h^{-1} \ {\rm kpc}
\end{equation}
based on Equations (13) and (14) at a \lya\ absorption equivalent width 
threshold $W = 0.3$ \AA.  The results apply over the $K$-band luminosity 
interval $0.005 \apl L_K \apl 3.3 L_{K_*}$ and for galaxies of all 
morphological types spanned by the observations.  

  To estimate the incidence and covering factor of tenuous gas in the extended 
halos, we perform a maximum-likelihood analysis with the probability that a 
galaxy gives rise to an absorption system of some absorption equivalent width 
threshold written as 
\begin{equation}
P=\epsilon \, H [r(L_B)-\rho]\kappa(\rho),
\end{equation}
where $\epsilon$ is the fraction of galaxies that give rise to \lya\ 
absorption, $H$ is the Heaviside step function that accounts for the scaling 
relationship between the gaseous extent of galaxies and galaxy $B$-band 
luminosity, and $\kappa$ is the covering factor of tenuous gas in the extended 
halos.  The likelihood of detecting an ensemble of galaxies, $n$ of which give 
rise to \lya\ absorption systems and $m$ of which do not, is given by
\begin{equation}
{\cal L}=\prod_{i=1}^n\,\epsilon H[r(L_{B_i})-\rho_i]\kappa(\rho_i)\times
\prod_{j=1}^m\,\{1-\epsilon H[r(L_{B_j})-\rho_j]\kappa(\rho_i)\}.
\end{equation}
Because it is difficult to separate $\epsilon$ from $\kappa$, we choose to 
simply determine the mean value of the product $\langle\epsilon\kappa\rangle$ 
as a result.  The likelihood function may therefore be written as
\begin{eqnarray}
{\cal L}&=& \prod_{i=1}^n\,\langle\epsilon\kappa\rangle\,\prod_{j=1}^m(1-\langle\epsilon\kappa\rangle) \\
        &=& \langle\epsilon\kappa\rangle^n\,(1-\langle\epsilon\kappa\rangle)^m.
\end{eqnarray}

  Based on the scaled $W$ versus $\rho$ anti-correlation presented in Figure 4,
we find that only two of the 31 galaxies at impact parameter $\rho < 180 \ 
h^{-1}$ kpc do not produce corresponding \lya\ absorption lines to within 
sensitive upper limits, while 11 of the 16 galaxies at impact parameter $\rho >
180 \ h^{-1}$ kpc do not produce corresponding \lya\ absorption lines to within
sensitive upper limits.  The maximum likelihood analysis yields a best estimate
of $\langle\epsilon\kappa\rangle = 0.94$ with a 1 $\sigma$ lower bound of 
$\langle\epsilon\kappa\rangle = 0.86$ for tenuous gas within a radius of 180 
kpc around the galaxies.

  We conclude that, at the rest-frame absorption equivalent width threshold
$W\apg 0.3$ \AA\ (corresponding to a neutral hydrogen column density threshold 
$N(\hI)\apg 1.3\times 10^{14}\,\cmjj$), most galaxies are surrounded by 
extended gaseous halos of $\approx 180 \ h^{-1}$ kpc radius with a covering 
factor of 94\%.  The agreement between Equations (17) and (20) demonstrates 
that the ground-based near-infrared photometry agrees very well with the 
space-based optical photometry and further supports that a typical $L_*$ 
galaxy, independent of morphological type, does indeed possess an extended 
gaseous halo of radius $\approx 180\ h^{-1}$ kpc.  

\subsection{Implications for The Evolution of Star Formation Rate Density}

  The scaling relationship between the extent of tenuous gas around galaxies 
and galaxy luminosity provides a means of quantitatively relating statistical 
properties of \lya\ absorption systems to statistical properties of faint 
galaxies.  We have pointed out in Paper I that given a known galaxy population 
and the known scaling relation we can estimate the fraction of \lya\ absorption
systems that originate in extended gaseous halos of galaxies.  But we have also
demonstrated that the prediction may vary from 30\% to 100\%.  The large 
uncertainty is primarily due to the uncertainties in the normalization and the 
faint-end slope of the galaxy luminosity function.  The result also reflects 
the generic difficulty in understanding galaxy formation and evolution based on
comparisons of observational quantities obtained from magnitude-limited galaxy 
surveys (which suffer from various selection biases) and theoretical 
predictions.  

  As discussed by Chen, Lanzetta, \& Fern\'andez-Soto (2000), the predicted 
number density of \lya\ absorption systems arising in the extended gaseous 
halos of galaxies may be given by
\begin{equation}
n(z) = \frac{c}{H_0} (1 + z) (1 + 2 q_0 z)^{-1/2} \int_0^\infty 
d\left(\frac{L_B}{L_{B_*}}\right) \ \Phi(L_B,z) \sigma(L_B) \kappa \epsilon,
\end{equation}
where $c$ is the speed of light, $\Phi(L_B,z)$ is the galaxy luminosity
function, $\sigma$ is the \hI\ absorbing gas cross section that scales with 
galaxy $B$-band luminosity, $\kappa$ is the halo covering factor, and 
$\epsilon$ is the fraction of galaxies that produce corresponding \lya\ 
absorption systems.  Substituting the scaling relationship according to 
Equations (15), (16), and (17), and adopting the result of \S\S\ 7.1 that most 
galaxies are surrounded by extended gaseous halos of $\approx 180 \ h^{-1}$ kpc
radius with a roughly unity covering factor, we find
\begin{equation}
n(z) = \frac{c}{H_0} (1 + z) (1 + 2 q_0 z)^{-1/2} \int_0^\infty 
d\left(\frac{L_B}{L_{B_*}}\right) \ \left(\frac{L_B}{L_{B_*}}\right)^{0.8}
\Phi(L_B,z) \pi r_*^2.
\end{equation}

  For comparison, the comoving $B$-band luminosity density is defined as
\begin{equation}
{\cal L}_B(z) = \int_0^\infty d\left(\frac{L_B}{L_{B_*}}\right)\,L_B\,\Phi(L_B,z).
\end{equation}
We can therefore relate the predicted number density of \lya\ absorption 
systems with ${\cal L}_B(z)$ by 
\begin{equation}
n(z) = \frac{c}{H_0} (1 + z) (1 + 2 q_0 z)^{-1/2} 
\left(\frac{\pi r_*^2}{L_{B_*}}\right)\,{\cal L}_B(z)\,{\cal O}(L_B),
\end{equation}
where ${\cal O}(L_B)$ accounts for the departure of the mean absorption gas 
cross section averaged over galaxies of different luminosities per unit 
comoving volume from the comoving $B$-band luminosity density and is defined as
\begin{equation}
{\cal O}(L_B) = 
\int_0^\infty\,d\left(\frac{L_B}{L_{B_*}}\right)\left(\frac{L_B}{L_{B_*}}\right)^{0.8}\,\Phi(L_B,z)\Bigg/
\int_0^\infty\,d\left(\frac{L_B}{L_{B_*}}\right)\left(\frac{L_B}{L_{B_*}}\right)\,\Phi(L_B,z).
\end{equation}
Performing these integrals, we find
\begin{equation}
{\cal O}(L_B)=\Gamma(1.8)/\Gamma(2.0)=0.93 \sim 1.0,
\end{equation}
where $\Gamma$ is the gamma function.

 It appears that the number density evolution of \lya\ absorption systems 
traces the evolution of the comoving galaxy $B$-band luminosity density.
Although the results of the analysis presented in \S\S\ 6.2 indicate that 
galaxy mass (as probed by galaxy $K$-band luminosity), rather than recent star
formation activity (as probed by galaxy $B$-band luminosity) is the dominant 
factor that determines the gaseous extent around galaxies, we argue that 
Equation (28) is valid so long as the strong correlation between galaxy 
$K$-band and $B$-band luminosities seen in our survey (Figure 7) remains for
galaxies at higher redshifts.

  Comparing the measurements of the \lya\ absorption line density obtained by
Weymann \etal\ (1998) at redshifts $z<1.5$ with the ones obtained by Bechtold 
(1994) at redshifts $z>1.5$ (see Chen \etal\ 2000), we find that the number 
density of \lya\ absorption systems increases gradually with redshift, implying
a steadily increasing comoving $B$-band luminosity density over the entire 
redshift range ($0\apl z\apl 4$).  The predicted shallow slope of the comoving 
$B$-band luminosity density as a function of redshift at redshifts $z<1$ (where
the scaling relation between the gaseous extent of galaxies and galaxy $B$-band
luminosity is well understood and measured) agrees better with the results 
presented by Cowie, Songaila, \& Barger (1999), but disagrees with the steep 
slope presented by Lilly \etal\ (1996).  On the other hand, it is in a broad 
agreement with recent measurements of luminosity density evolution at higher 
redshifts (Pascarelle \etal\ 1998; Steidel \etal\ 1999).  Namely, the comoving 
galaxy luminosity density does not fall off at redshifts beyond $z=3$.

  An accurate assessment of luminosity density evolution bears importantly on
discriminating between different galaxy formation scenarios (e.g.\ Somerville 
\& Primack 1998).  A flat or steadily increasing galaxy luminosity density 
would imply that the bulk of star formation occurs much earlier (c.f.\ Madau, 
Pozzetti, \& Dickinson 1998), which would present a serious challenge to the
hierarchical formation model.  We have demonstrated that if (1) all the 
observed \lya\ absorption systems of neutral hydrogen column density $N(\hI) 
\apg 10^{14}$ \cmjj\ arise in extended gaseous halos around galaxies and (2)
the scaling relation applies to absorbers at all redshifts, then the number 
density evolution of \lya\ absorption systems may serve to constrain the 
measurements of cosmic star formation rate density.

\section{SUMMARY AND CONCLUSIONS}

  We present complete results of a program to obtain and analyze HST WFPC2 
images and ground-based $K'$ images of galaxies identified in an imaging and 
spectroscopic survey of faint galaxies in fields of HST spectroscopic target 
QSOs.  We measure properties of 142 galaxies, of which 40 are associated with 
corresponding \lya\ absorption systems and 47 do not produce corresponding
\lya\ absorption lines to within sensitive upper limits.  The galaxy and
absorber pair sample is about 50\% larger than the one previously published in
Paper I.  We repeat the likelihood analysis to examine whether or not the 
results of previous analysis may be retained and further improved by including 
additional measurements.

  Following Paper I, we consider galaxy and absorber pairs physically 
correlated if (1) the galaxy--absorber cross-correlation amplitude satisfies 
$\xi_{\rm ga}(v,\rho) > 1$ and (2) the galaxy impact parameter satisfies $\rho 
< 200 \ h^{-1}$ kpc.  Including only galaxy and absorber pairs that are likely 
to be physically associated and excluding galaxy and absorber pairs within 3000
\kms\ of the background QSOs leaves 34 galaxy and absorber pairs and 13 
galaxies that do not produce corresponding \lya\ absorption lines to within 
sensitive upper limits.  Redshifts of the galaxy and absorber pairs range from 
$z = 0.0752$ to 0.8920 with a median of $z = 0.3567$, and impact parameter 
separations of the galaxy and absorber pairs range from $\rho = 12.4$ to $175.2
\ h^{-1}$ kpc with a median of $\rho = 62.2 \ h^{-1}$ kpc.  Of the 47 galaxies,
15 (32\%) are of $B$-band luminosity $L_B < 0.25 \ L_{B_*}$ with redshifts 
ranging from $z = 0.0915$ to $z = 0.6350$, and six (13\%) are of low surface 
brightness with redshifts ranging from $z = 0.0915$ to $z = 0.3180$.

  We confirm the results previously published in Paper I with improved 
statistics that the amount of gas encountered along the line of sight depends 
on the galaxy impact parameter $\rho$, galaxy $B$-band luminosity $L_B$, but 
does not depend strongly on the galaxy mean surface brightness $\langle \mu 
\rangle$, disk-to-bulge ratio $D/B$, or redshift $z$ and that there is no 
evidence that galaxy interactions play an important role in distributing 
tenuous gas around galaxies in most cases.  Furthermore, we find that:

  1.  The anti-correlation between \lya\ absorption equivalent width $W$ and
galaxy impact parameter $\rho$ accounting for galaxy $K$-band luminosity $L_K$
is superior to the fiducial relationship between $W$ and $\rho$.  We conclude
that the amount of gas intercepted along the line of sight depends on galaxy
$K$-band luminosity which, together with the fact that galaxies of all 
morphological types possess extended gaseous halos, indicates that galaxy mass
is likely to be the dominant factor that determines the extent of tenuous gas 
around galaxies. 

  2.  The relationship between galaxy gaseous radius $r$ and galaxy $K$-band
luminosity $L_K$ may be described by
\begin{equation}
\frac{r}{r_*} = \left( \frac{L_K}{L_{K_*}} \right)^{t_K},
\end{equation}
with
\begin{equation}
t_K = 0.28 \pm 0.08
\end{equation}
and 
\begin{equation}
r_* = 177_{-26}^{+30} \ h^{-1} \ {\rm kpc}
\end{equation}
for a \lya\ absorption equivalent width threshold $W = 0.3$ \AA.  The $t=0$ 
case (no dependence of gaseous radius on galaxy $K$-band luminosity) can be 
ruled out at the $3.5 \sigma$ level of significance.  

  3.  At the rest-frame absorption equivalent width threshold $W\apg 0.3$ 
\AA\ (which corresponds to a neutral hydrogen column density threshold 
$N(\hI)\apg 1.3\times 10^{14}\,\cmjj$), we find that a typical $L_*$ galaxy is
surrounded by an extended gaseous halo of $\approx 180 \ h^{-1}$ kpc radius 
with a covering factor of 94\%.

  4.  Adopting the scaling relationship between the extent of tenuous gas 
around galaxies and galaxy $B$-band luminosity, we find that the predicted 
number density of \lya\ absorption systems arising in extended gaseous halos of
galaxies is equivalent to the comoving $B$-band luminosity density.  Therefore,
the observed number density evolution of \lya\ absorption systems may serve to 
constrain the measurements of cosmic star formation rate density.

\acknowledgments

  The authors thank the staff of STScI for their expert assistance and Sam 
Pascarelle for helping to obtain some of the near-infrared images.  This work 
was supported by STScI grant GO--07290.01--96A and NSF grant AST--9624216.

\newpage

\figcaption{Final images of galaxies obtained with HST using WFPC2 with the 
F702W.  The spatial extent of each image is roughly 25 $h^{-1}$ kpc on a side,
and the orientation of each image is arbitrary.}

\figcaption{Residuals of the $W$ vs. $\rho$ anti-correlation as a function of 
galaxy $B$-band luminosity $L_B$ (the upper-left panel), redshift $z$ (the
upper-right panel), galaxy mean surface brightness $\langle \mu \rangle$ 
(the lower-left panel), and galaxy disk-to-bulge ratio $D/B$ (the lower-right 
panel).  Circles represent elliptical or S0 galaxies, triangles represent 
early-type spiral galaxies, and squares represent late-type spiral galaxies. 
Points with arrows indicate $3 \sigma$ upper limits.  The dotted line in the 
lower-left panel at $\langle\mu\rangle = 25.7$ indicates the borderline between
low surface brightness and high surface brightness galaxies.}

\figcaption{Logarithm of \lya\ rest-frame equivalent width $W$ vs.\ logarithm 
of galaxy impact parameter $\rho$.  Symbols are the same as those in Figure 2.
The cosmic scatter is indicated by the error bar in the upper-right corner.}

\figcaption{Logarithm of \lya\ rest-frame equivalent width $W$ vs.\ logarithm 
of galaxy impact parameter $\rho$ scaled by galaxy $B$-band luminosity. The 
scaling factor is determined from the analysis described in \S\S\ 6.1.  Symbols
are the same as those in Figure 2.  The cosmic scatter is indicated by the 
error bar in the upper-right corner.}

\figcaption{Logarithm of neutral hydrogen column density $N$ vs.\ logarithm of 
galaxy impact parameter $\rho$ scaled by galaxy $B$-band luminosity.  The 
scaling factor is determined from the analysis described in \S\S\ 6.1.  
Neutral hydrogen column densities are determined from \lya\ rest-frame 
equivalent widths under the assumption that Doppler parameters are contained in
the range $20 < b < 40$ \kms.  Symbols are the same as those in Figure 2, and 
the cosmic scatter is indicated by the error bar in the upper-right corner.}

\figcaption{Comparison of the $W$ vs.\ $\rho$ anti-correlation scaled by galaxy
$K$-band luminosity (the left panel) and by galaxy $B$-band luminosity (the 
right panel).  The scaling factor is determined from the analysis described in 
\S\S\ 6.1.  Symbols are the same as those in Figure 2, and the cosmic scatter 
is indicated by the error bar in the upper-right corner.}

\figcaption{Galaxy $K$-band luminosity vs.\ galaxy $B$-band luminosity for
galaxies in the sample with $K$-band photometric measurements available.
Symbols are the same as those in Figure 2.}





\newpage

\plotone{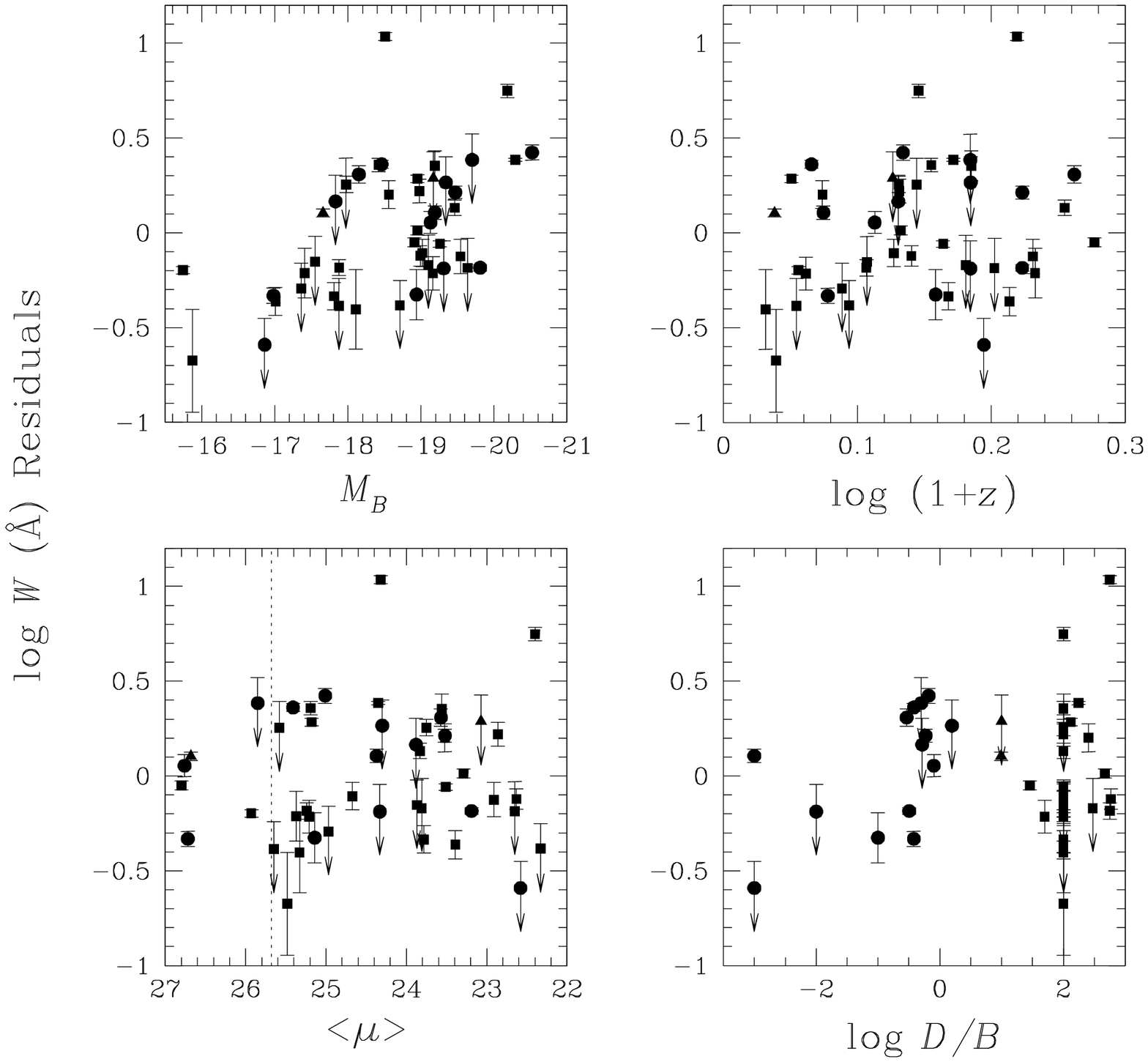}

\newpage

\plotone{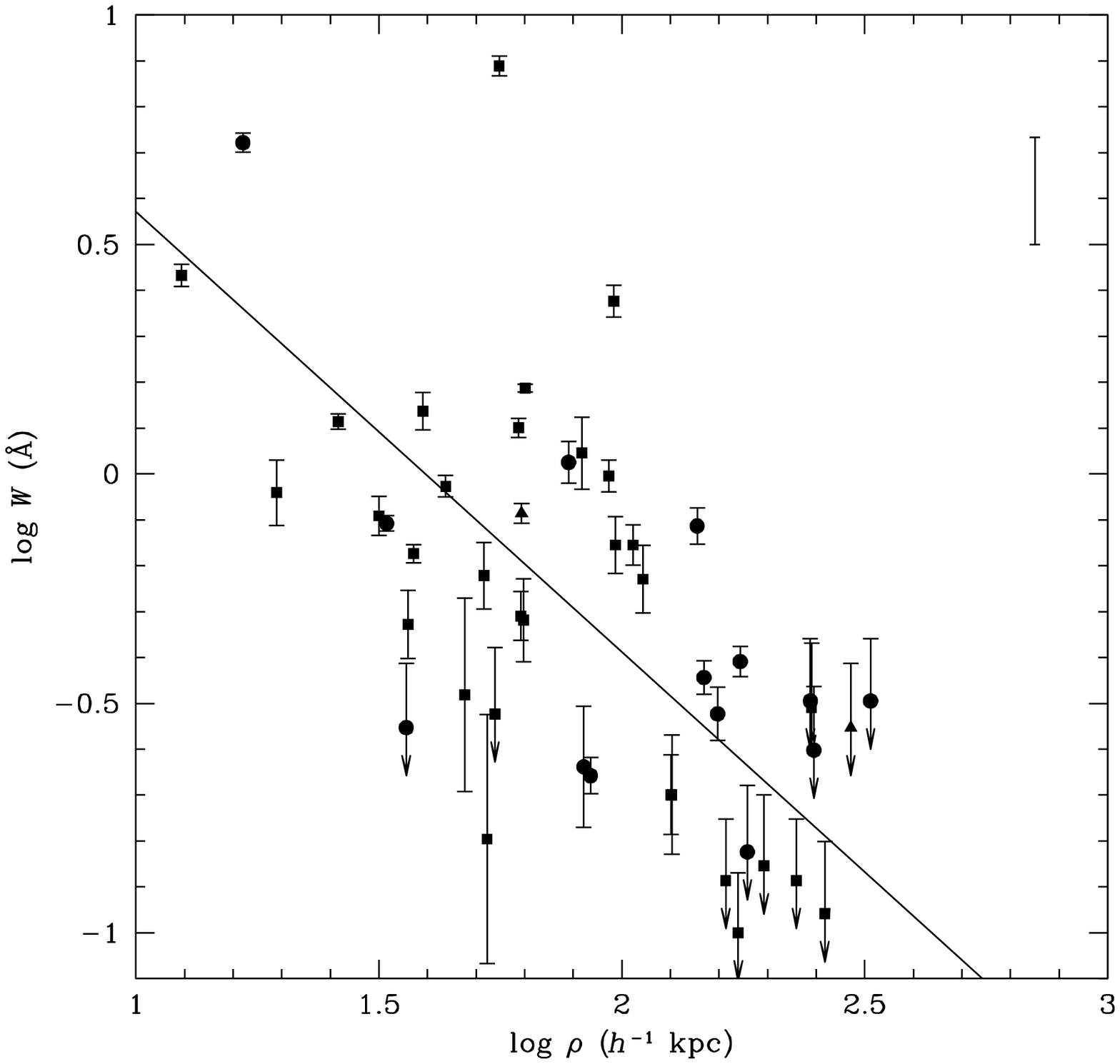}

\newpage

\plotone{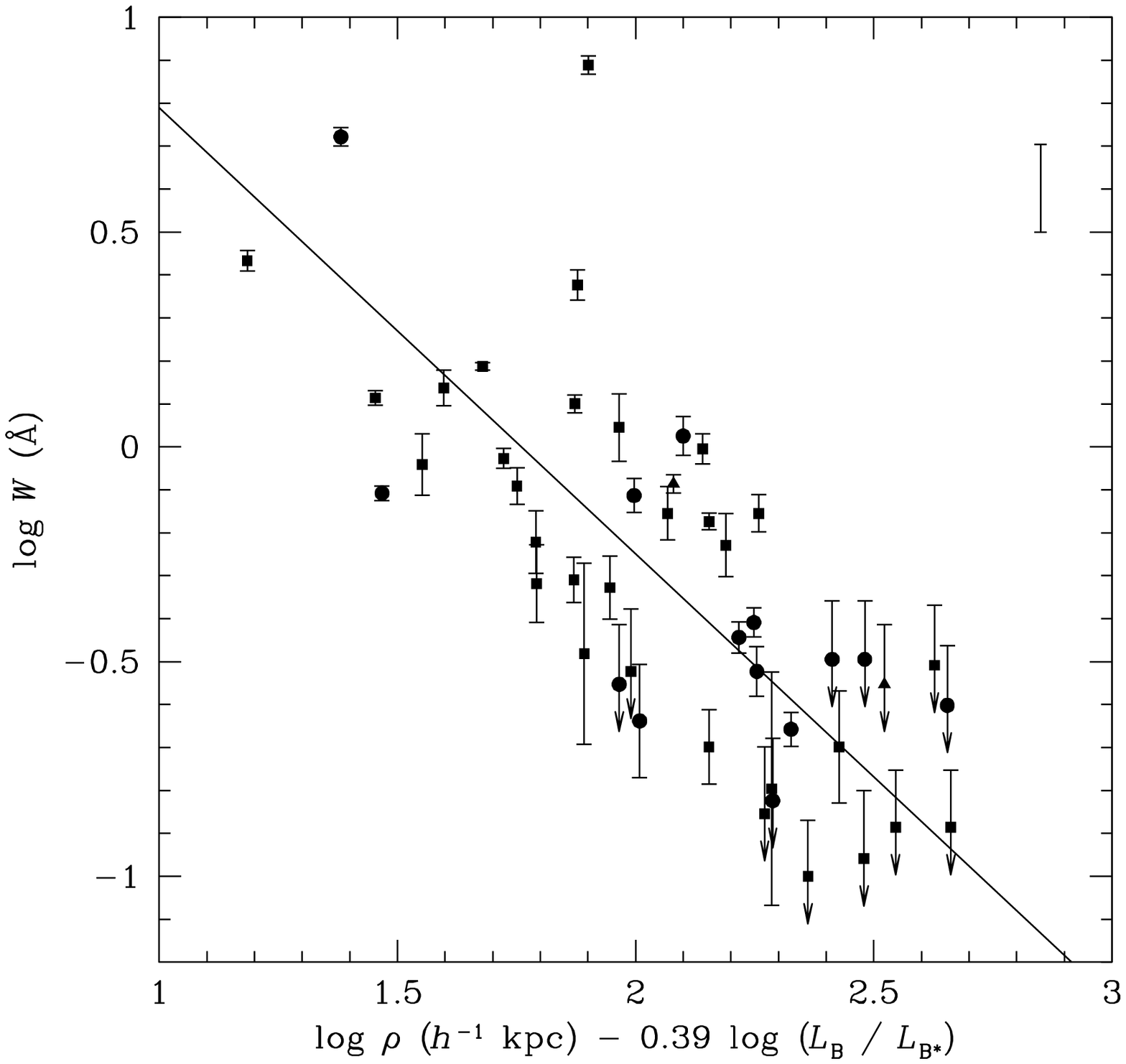}

\newpage

\plotone{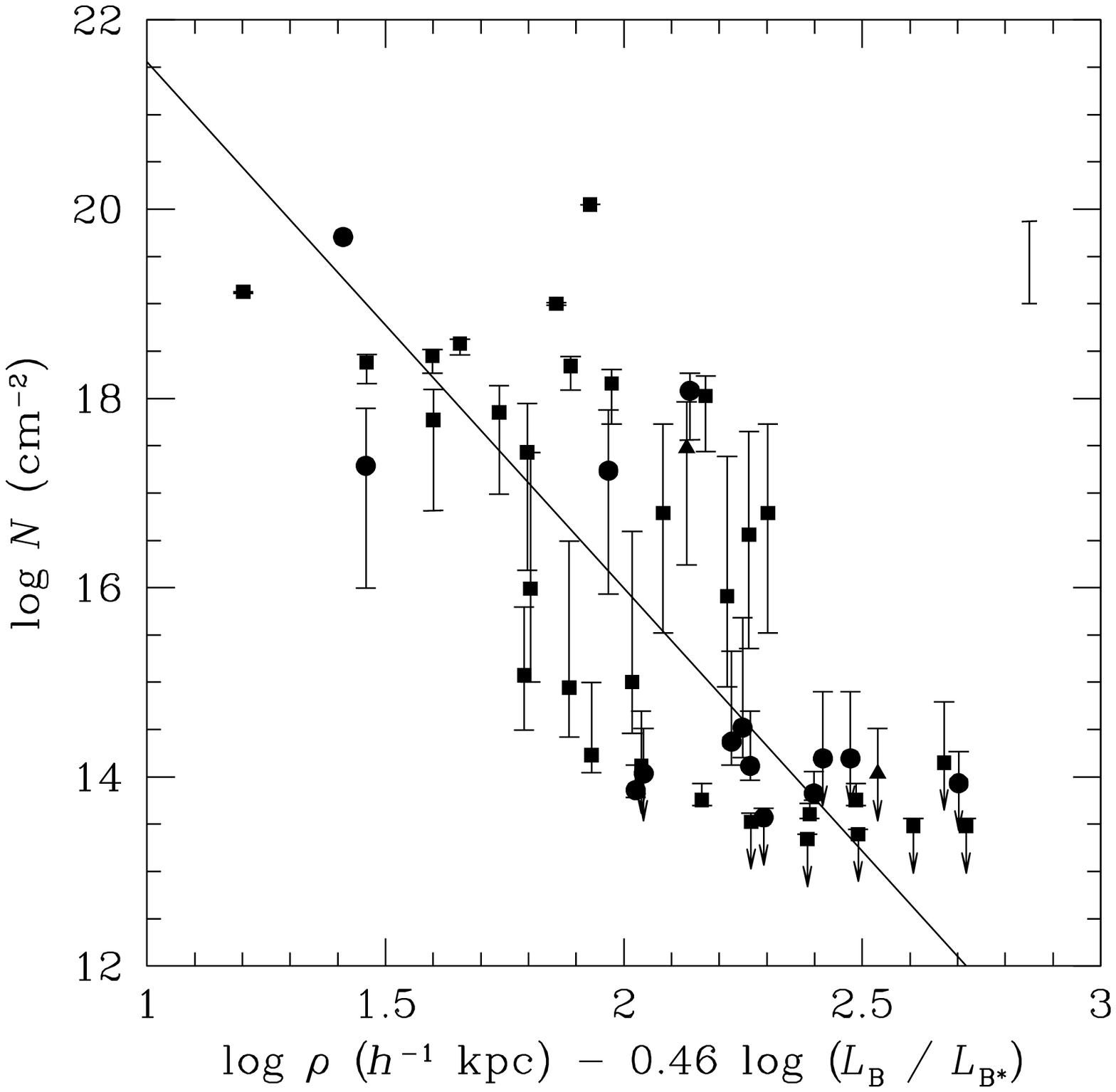}

\newpage

\plotone{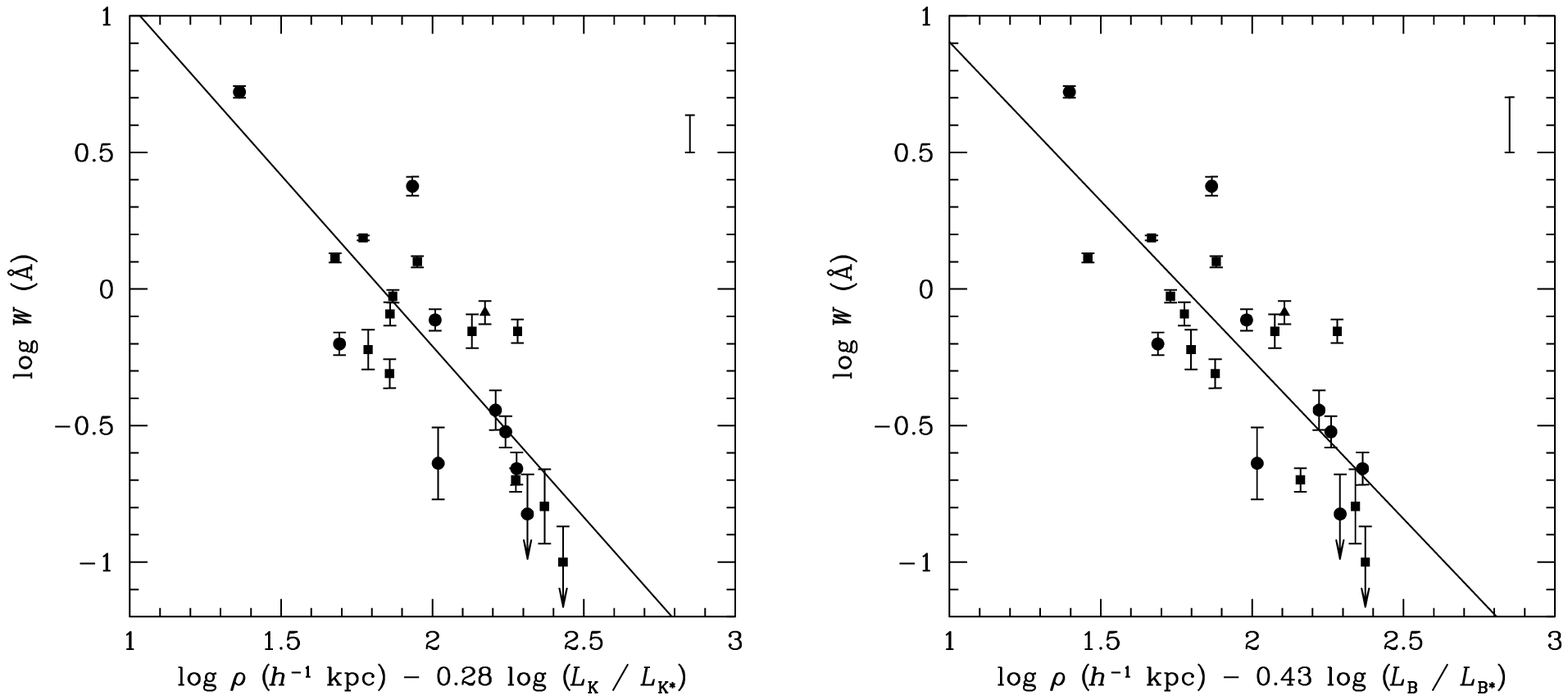}

\newpage

\plotone{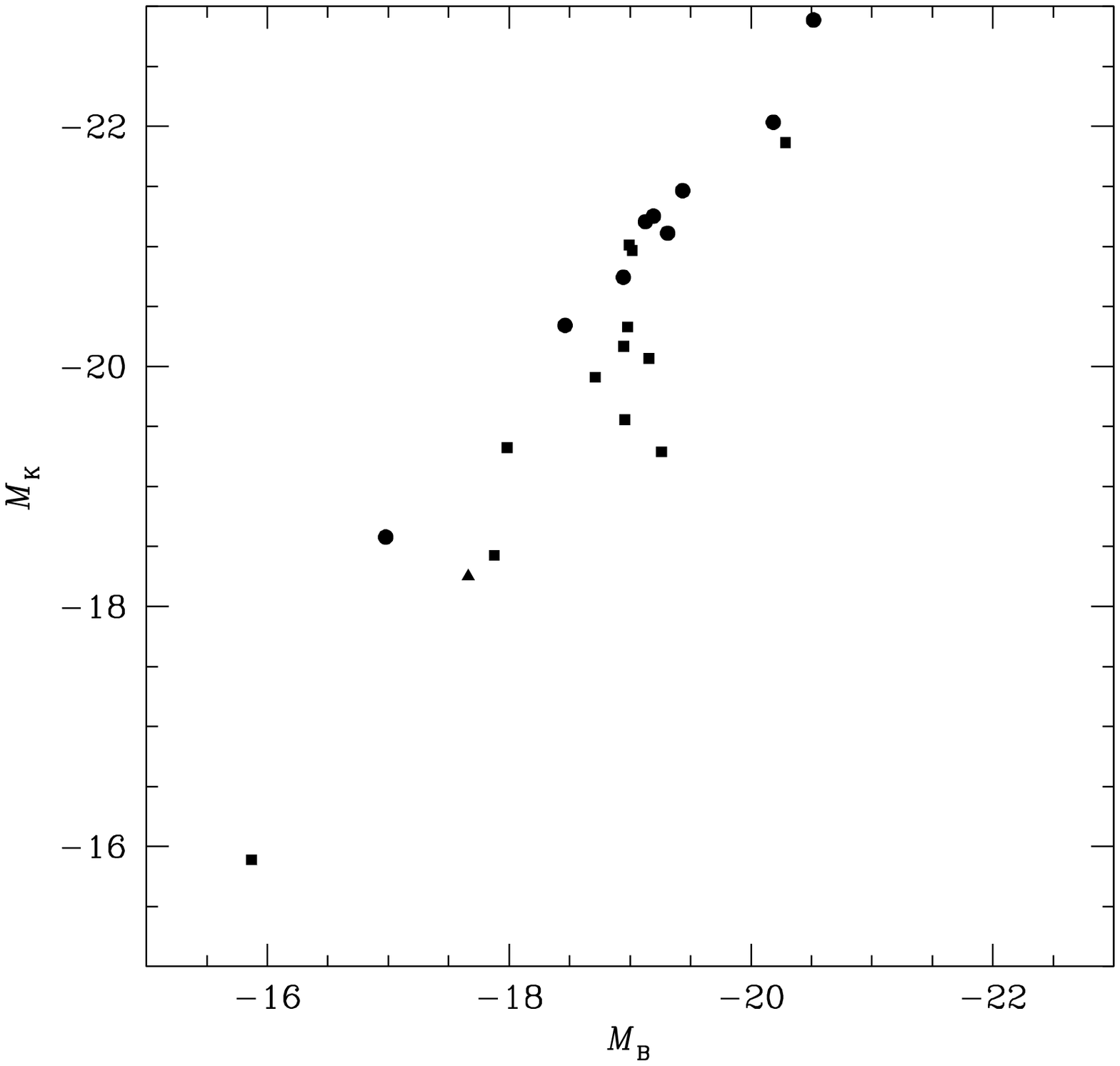}

\newpage

\begin{table}
\begin{center}
\begin{tabular}{p{1.5in}cccccr}
\multicolumn{7}{c}{Table 1} \\
\multicolumn{7}{c}{Journal of Observations} \\
\hline
\hline
& & & & & Exposure \\
\multicolumn{1}{c}{Field} & $\alpha$ (2000) & $\delta$ (2000) & $z_{\rm em}$ &
 Filter & Time (s) & \multicolumn{1}{c}{Date} \\
\tableline
0122$-$0021 \dotfill & 01:25:28.8 & $-$00:05:55.8 & 1.070 & F702W & 2100 &  3
Dec 1997\\
0405$-$1219 \dotfill & 04:07:48.4 & $-$12:11:36.0 & 0.574 & F702W & 2100 & 16
Sep 1998\\
0405$-$1219 \dotfill & 04:07:48.4 & $-$12:11:36.0 & 0.574 & F702W & 2100 & 23
Sep 1998\\
0903$+$1658 \dotfill & 09:06:31.9 & $+$16:46:11.5 & 0.412 & F702W & 2100 & 16
Nov 1997\\
1136$-$1334 \dotfill & 11:39:10.7 & $-$13:50:43.5 & 0.560 & F702W & 2100 & 10
Jun 1997\\
1216$+$0657 \dotfill & 12:19:20.9 & $+$06:38:38.4 & 0.331 & F702W & 2100 & 30
Mar 1998\\
1259$+$5920 \dotfill & 13:01:12.9 & $+$59:02:06.1 & 0.478 & F702W & 2100 & 20
Jan 1997\\
1259$+$5920 \dotfill & 13:01:12.9 & $+$59:02:06.1 & 0.478 & F702W & 2100 & 25
Oct 1998\\
1424$-$1150 \dotfill & 14:27:38.2 & $-$12:03:50.5 & 0.806 & F702W & 2100 &  7
Feb 1997\\
1641$+$3954 \dotfill & 16:42:58.7 & $+$39:48:36.0 & 0.593 & F702W & 2100 &  8
Sep 1998\\
2251$+$1552 \dotfill & 22:53:57.7 & $+$16:08:53.6 & 0.859 & F702W & 2100 &  6
Dec 1997\\
\hline
\end{tabular}
\end{center}
\end{table}

\newpage

\begin{table}
\begin{center}
\begin{tabular}{p{1.5in}cccccr}
\multicolumn{7}{c}{Table 2} \\
\multicolumn{7}{c}{Journal of Archival Observations} \\
\hline
\hline
& & & & & Exposure \\
\multicolumn{1}{c}{Field} & $\alpha$ (2000) & $\delta$ (2000) & $z_{\rm em}$ &
 Filter & Time (s) & \multicolumn{1}{c}{Date} \\
\tableline
1317$+$2743 \dotfill & 13:19:56.3 & $+$27:28:08.4 & 1.022 & F702W & 4700 &  1
Jun 1995\\
\hline
\end{tabular}
\end{center}
\end{table}

\newpage

\begin{table}
\begin{center}
\begin{tabular}{p{1.5in}ccccc}
\multicolumn{6}{c}{Table 3} \\
\multicolumn{6}{c}{Summary of Other Observations} \\
\hline
\hline
& \multicolumn{2}{c}{Galaxies} & & \multicolumn{2}{c}{Absorbers} \\
\cline{2-3}
\cline{5-6}
& Number & & & Number & \\
\multicolumn{1}{c}{Field} & Included & Reference & &
Included & Reference \\
\hline
0122$-$0021 \dotfill &  4 &   1 & &  3 & 1,2,3 \\
0405$-$1219 \dotfill & 15 & 1,4 & &  5 &     1 \\
0903$+$1658 \dotfill & 13 & 1,4 & &  0 &     1 \\
1136$-$1334 \dotfill & 13 &   1 & &  0 &   1,2 \\
1216$+$0657 \dotfill &  5 &   1 & &  1 &     1 \\
1259$+$5920 \dotfill &  4 &   1 & &  1 &   1,2 \\
1317$+$2743 \dotfill &  4 &   5 & &  2 &   1,2 \\
1424$-$1150 \dotfill &  7 &   1 & &  1 &   1,2 \\
1641$+$3954 \dotfill & 15 &   4 & &  1 &     1 \\
2251$+$1552 \dotfill &  1 &   1 & &  1 &   1,2 \\
\hline
\end{tabular}
\parbox{5.0 in}{\hspace{0.25 in} REFERENCES---(1) our own observations and 
analysis, in preparation; (2) Bahcall et al.\ 1993; (3) Bahcall et al.\ 1996; 
(4) Ellingson \& Yee 1994; (5) Le Brun, Bergeron, \& Boiss\'e 1996.}
\end{center}
\end{table}

\end{document}